\documentclass[epj,final]{svjour}

\usepackage{latexsym}
\usepackage{graphics}
\usepackage{epsfig}
\usepackage{rotate}
\usepackage{amsmath}
\usepackage{amssymb}
\usepackage{array}
\usepackage[dvips]{color}
\usepackage{afterpage}
\usepackage{graphics}

\newcommand{\GeV}{\mathrm{GeV}}
\newcommand{\MeV}{\mathrm{MeV}}
\newcommand{\cm}{\mathrm{cm}}

\newcommand{\tcs}{\sigma}
\newcommand{\tdcs}{d\sigma/dQ^2}
\newcommand{\tlB}{\begin{array}{r} \tcs \\ \tdcs \end{array}}
\newcommand{\tlC}{\begin{array}{r} \tcs \\ \tdcs \\ \tcs \otimes \tdcs \end{array}}
\newcommand{\tlD}{\tcs \otimes \tdcs}
\newcommand{\tnA}[1]{\begin{array}{r} #1 \end{array}}
\newcommand{\tnB}[2]{\begin{array}{r} #1 \\ #2 \end{array}}
\newcommand{\tnC}[3]{\begin{array}{r} #1 \\ #2 \\ #3 \end{array}}

\newcommand{\cs}[1]{\langle\sigma_{#1}\rangle}
\newcommand{\dcs}[1]{\langle\tilde{\sigma}_{#1}\rangle}

\begin{document}
\hugehead

\title{A study of quasi-elastic muon neutrino and antineutrino scattering in the NOMAD experiment}

\author{
V.~Lyubushkin\inst{6} \and
B.~Popov\inst{6,14} \and
J.J.~Kim\inst{19} \and 
L.~Camilleri\inst{8} \and
J.-M.~Levy\inst{14} \and
M.~Mezzetto\inst{13} \and
D.~Naumov\inst{6,7} \and 
S.~Alekhin\inst{25} \and
P.~Astier\inst{14} \and 
D.~Autiero\inst{8}
% ...
}

\mail{Vladimir.Lyubushkin@cern.ch}

\institute{LAPP, Annecy, France \and 
Johns Hopkins Univ., Baltimore, MD, USA \and
Harvard Univ., Cambridge, MA, USA \and
Univ. of Calabria and INFN, Cosenza, Italy \and
Dortmund Univ., Dortmund, Germany \and
JINR, Dubna, Russia \and
Univ. of Florence and INFN,  Florence, Italy \and
CERN, Geneva, Switzerland \and
University of Lausanne, Lausanne, Switzerland \and
UCLA, Los Angeles, CA, USA \and
University of Melbourne, Melbourne, Australia \and
Inst. for Nuclear Research, INR Moscow, Russia \and
Univ. of Padova and INFN, Padova, Italy \and
LPNHE, Univ. of Paris VI and VII, Paris, France \and
Univ. of Pavia and INFN, Pavia, Italy \and
Univ. of Pisa and INFN, Pisa, Italy \and
Roma Tre University and INFN, Rome, Italy \and
DAPNIA, CEA Saclay, France \and
Univ. of South Carolina, Columbia, SC, USA \and
Univ. of Sydney, Sydney, Australia \and
Univ. of Urbino, Urbino, and INFN Florence, Italy \and
IFIC, Valencia, Spain \and
Rudjer Bo\v{s}kovi\'{c} Institute, Zagreb, Croatia \and
ETH Z\`urich, Z\`urich, Switzerland \and
Inst. for High Energy Physics, 142281, Protvino, Moscow region, Russia
}

\date{Received: date / Revised version: \today}

\abstract{ 
We have studied the muon neutrino and antineutrino quasi-elastic (QEL) scattering
reactions ($\nu_{\mu} n\to \mu^{-} p$ and $\bar{\nu}_{\mu} p\to \mu^{+}n$)
using a set of experimental data collected by the NOMAD collaboration. 
We have performed measurements of the cross-section of these processes on a nuclear
target (mainly Carbon) normalizing it to the 
total $\nu_{\mu}$ ($\bar{\nu}_{\mu}$) charged current cross-section.
The results for the flux averaged QEL cross-sections in 
the (anti)neutrino energy interval $3-100~\GeV$ are 
$\cs{qel}_{\nu_\mu} = (0.92 \pm 0.02 (stat) \pm 0.06 (syst))\times 10^{-38}~\cm^2$
and 
$\cs{qel}_{\bar{\nu}_\mu} = (0.81 \pm 0.05 (stat) \pm 0.08 (syst)) \times 10^{-38}~\cm^2$
for neutrino and antineutrino, respectively.
The axial mass parameter $M_A$ was extracted from the measured quasi-elastic neutrino
cross-section. The corresponding result 
is $M_A = 1.05 \pm 0.02 (stat) \pm 0.06 (syst)~\GeV$. It is consistent with 
the axial mass values recalculated from the antineutrino cross-section 
and extracted from the pure $Q^2$ shape analysis of the high purity sample of $\nu_\mu$ 
quasi-elastic 2-track events, but has smaller systematic error and should be quoted 
as the main result of this work. Our measured $M_A$ is found to be in
good agreement with the world average value obtained in previous deuterium filled bubble 
chamber experiments. 
The NOMAD measurement of $M_A$ is lower than 
those recently published by K2K and MiniBooNE collaborations.
However, within the large errors quoted by these experiments
on $M_A$,  these results are compatible with the more precise NOMAD value.
\PACS{
      {13.15.+g}{Neutrino interactions}       \and
      {25.30.Pt}{Neutrino-induced reactions}  
     }
\keywords{neutrino interactions, neutrino and antineutrino quasi-elastic scattering}
}

\maketitle

\section{Introduction}

A precise knowledge of the cross-section of (anti)neutrino-nucleus quasi-elastic 
scattering process (QEL) is  important for the planning and analysis of any experiment 
which detects astrophysical, atmospheric or accelerator neutrinos. 
The available measurements from early experiments at 
ANL~\cite{Kustom:1969dh,Mann:1973pr,Barish:1977qk,Miller:1982qi},
BNL~\cite{Fanourakis:1980si,Baker:1981su,Ahrens:1988rr,Kitagaki:1990vs},
FNAL~\cite{Kitagaki:1983px,Asratian:1984ir},
CERN~\cite{Block:1964gj,Orkin-Lecourtois:1967,Holder:1968,Budagov:1969bg,Bonetti:1977cs,Armenise:1979zg,Pohl:1979zm,Allasia:1990uy}
and IHEP\cite{Belikov:1981ut,Belikov:1983kg,Grabosch:1986js,Brunner:1989kw}
have considerable errors due to low statistics and a lack of knowledge of the 
precise incoming neutrino flux. Unfortunately, even within these large errors, the results 
are often conflicting.

This subject remains very topical. Recently several attempts have been
made to investigate the QEL process in the data collected by modern accelerator neutrino
experiments (such as NuTeV~\cite{Suwonjandee:04},
K2K~\cite{Gran:2006jn,Mariani:2008zz} and MiniBooNE~\cite{MiniBooNE:2007ru}). 
Unfortunately they have not clarified the situation again due to large errors 
assigned to their measurements. Dedicated experiments, such as e.g. 
SciBooNE~\cite{AguilarArevalo:2006se} and MINER$\nu$A~\cite{Drakoulakos:2004gn}, 
are now being performed.

In the present analysis, we study both $\nu_\mu$ and $\bar{\nu}_\mu$ QEL scattering 
in the data collected by the NOMAD collaboration. The NOMAD detector was 
exposed to a wide-band neutrino beam produced by the 450~GeV proton synchrotron 
(SPS, CERN). A detailed description of the experimental set-up can be found 
in~\cite{Altegoer:1997gv}. The characteristics of the incoming neutrino flux 
are given in~\cite{Astier:2003rj}.

The large amount of collected data and the good quality of event reconstruction 
in the NOMAD detector provide a unique possibility to measure the QEL cross-section 
with a combination of
small statistical and systematic errors. The data sample used in this 
analysis consists of about 751000\,(23000) $\nu_\mu$ ($\bar{\nu}_\mu$) 
charged-current (CC) interactions in a reduced detector fiducial volume.
The average energy of the incoming $\nu_\mu$ ($\bar{\nu}_\mu$) is 25.9 (17.6)~GeV.

The merit of the current analysis is the possibility of keeping 
the systematic error relatively small. It takes advantage 
from three main factors: 1) the NOMAD detector is capable of selecting
a sample of QEL events with a high purity and a good efficiency; 
2) a simultaneous measurement of both two track and single track 
$\nu_\mu$ QEL events allows to constrain the systematics associated
with nuclear reinteractions; 3) a wide energy range of the NOMAD neutrino 
beam allows to perform a precise normalization to the well-known total 
(DIS) $\nu_\mu$ CC cross-section.

The paper is organized as follows. 
In Section~\ref{section:exp_data} we give a brief review of the published 
experimental data on QEL (anti)neutrino scattering. The NOMAD detector and 
the incoming neutrino flux are briefly discussed in Section~\ref{section:nomad}. 
In Section~\ref{section:nomad_mc} we outline the MC modeling of signal and 
background events, emphasizing also the importance of nuclear effects. 
Section~\ref{section:QEL_selection} is devoted to the selection of the QEL events; 
we describe the QEL identification procedure and compare 
the MC predictions with experimental data. The methods used to measure 
the QEL cross-section and the phenomenological axial mass parameter $M_A$ are 
the subjects of Section~\ref{section:QEL_measurement}. The systematic uncertainties 
are summarized in Section~\ref{section:systematics}. The results are presented in
Section~\ref{section:results}. Finally, a summary and discussion of the obtained 
results are given in Section~\ref{section:discussion}.
 
More detailed information can be found in~\cite{Lyubushkin:2008zz}.

\newpage

\section{Review of existing experimental data on Quasi-Elastic (anti)neutrino scattering}
\label{section:exp_data}

\begin{table*}[htb]
\caption{\label{tab:ma_numu}
A summary of existing experimental data: the axial mass $M_A$ as measured in previous neutrino experiments. Numbers of observed events have been taken from the original papers; usually they are not corrected for efficiency and purity (the so-called QEL candidates). The axial mass value for the NuTeV experiment~\cite{Suwonjandee:04} was estimated from the published neutrino quasi-elastic cross section ($\sigma^{qel}_\nu=(0.94\pm0.03(stat)\pm0.07(syst))\times 10^{-38}\,\cm^2$); the systematic error for IHEP SKAT 90~\cite{Brunner:1989kw} is $0.14\,\GeV$. \hfill
}
\begin{tabular}{>{\em}l >{\em}l r >{$}r<{$}@{\hspace{5pt}}>{$}r<{$} >{\hspace{0pt}}c}
\hline\noalign{\smallskip}
\multicolumn{1}{l}{Experiment} & \multicolumn{1}{l}{Target} &
Events & \multicolumn{1}{r}{Method} & \multicolumn{1}{r}{$M_A,\,\GeV$} & {Ref.} \\
\noalign{\smallskip}\hline\noalign{\smallskip}
ANL 69      & Steel     &       & \tdcs & \tnA{1.05 \pm 0.20}
                                & \cite{Kustom:1969dh} \\
ANL 73      & Deuterium & 166   & \tlC & \tnC{0.97 \pm 0.16}{0.94 \pm 0.18}{0.95 \pm 0.12}
                                & \cite{Mann:1973pr} \\
ANL 77      & Deuterium & $\sim 600$ & \tlC
                                & \tnC{0.75^{+0.13}_{-0.11}}{1.01 \pm 0.09}{0.95 \pm 0.09}
                                & \cite{Barish:1977qk} \\
ANL 82      & Deuterium & 1737  & \tlC & \tnC{0.74 \pm 0.12}{1.05 \pm 0.05}{1.03 \pm 0.05}
                                & \cite{Miller:1982qi} \\
\noalign{\smallskip}\hline\noalign{\smallskip}
BNL 81      & Deuterium & 1138  & \tdcs & \tnA{1.07 \pm 0.06}
                                & \cite{Baker:1981su} \\
BNL 90      & Deuterium & 2538  & \tdcs & \tnA{1.070^{+0.040}_{-0.045}}
                                & \cite{Kitagaki:1990vs} \\
\noalign{\smallskip}\hline\noalign{\smallskip}
FermiLab 83 & Deuterium & 362   & \tdcs & \tnA{1.05^{+0.12}_{-0.16}}
                                & \cite{Kitagaki:1983px} \\
NuTeV 04    & Steel     & 21614 & \tcs  & \tnA{1.11 \pm 0.08} 
                                & \cite{Suwonjandee:04} \\
MiniBooNE 07 & Mineral oil & 193709 & \tdcs & \tnA{1.23 \pm 0.20}
                                & \cite{MiniBooNE:2007ru} \\
\noalign{\smallskip}\hline\noalign{\smallskip}
CERN HLBC 64 & Freon    & 236   & \tdcs & \tnA{1.00^{+0.35}_{-0.20}}
                                & \cite{Block:1964gj} \\
CERN HLBC 67 & Freon    &  90   & \tlD & \tnA{0.75^{+0.24}_{-0.20}}
                                & \cite{Orkin-Lecourtois:1967} \\
CERN SC 68   & Steel    & 236   & \tdcs & \tnA{0.65^{+0.45}_{-0.40}}
                                & \cite{Holder:1968} \\
CERN HLBC 69 & Propane  & 130   & \tlD & \tnA{0.70 \pm 0.20}
                                & \cite{Budagov:1969bg} \\
CERN GGM 77  & Freon    & 687   & \tlB & \tnB{0.88 \pm 0.19}{0.96 \pm 0.16}
                                & \cite{Bonetti:1977cs} \\
CERN GGM 79  & Propane/Freon & 556 & \tlB & \tnB{0.87 \pm 0.18}{0.99 \pm 0.12}
                                & \cite{Pohl:1979zm} \\
CERN BEBC 90 & Deuterium & 552  & \tlB & \tnB{0.94 \pm 0.07}{1.08 \pm 0.08}
                                & \cite{Allasia:1990uy} \\
\noalign{\smallskip}\hline\noalign{\smallskip}
IHEP 82      & Aluminium & 898  & \tdcs & \tnA{1.00 \pm 0.07}
                                & \cite{Belikov:1981ut} \\
IHEP 85      & Aluminium & 1753 & d\sigma_{\nu+\bar{\nu}}/dQ^2 & \tnA{1.00 \pm 0.04}
                                & \cite{Belikov:1983kg} \\
IHEP SCAT 88 & Freon     &  464 & \tlD & \tnA{0.96 \pm 0.15}
                                & \cite{Grabosch:1986js} \\ 
IHEP SCAT 90 & Freon     &      & \tlC & \tnC{1.08 \pm 0.07}{1.05 \pm 0.07}{1.06 \pm 0.05}
                                & \cite{Brunner:1989kw} \\
\noalign{\smallskip}\hline\noalign{\smallskip}
K2K 06, SciFi & Water    & $\sim 12000$ & \tdcs & \tnA{1.20 \pm 0.12}
                                & \cite{Gran:2006jn}\\
K2K 08, SciBar & Carbon  &      & \tdcs & \tnA{1.144 \pm 0.077}
                                & \cite{Mariani:2008zz}\\
\noalign{\smallskip}\hline
\end{tabular}
\vspace{10pt}
\caption{\label{tab:ma_anumu}
The same as in Table~\ref{tab:ma_numu}, but for antineutrino experiments. The axial mass value for the NuTeV experiment~\cite{Suwonjandee:04} was estimated from the published antineutrino quasi-elastic cross section ($\sigma^{qel}_{\bar{\nu}}=(1.12\pm0.04(stat)\pm0.10(syst))\times 10^{-38}\,\cm^2$); the systematic error for IHEP SKAT 90~\cite{Brunner:1989kw} is $0.20\,\GeV$. \hfill
}
\begin{tabular}{>{\em}l >{\em}l r >{$}r<{$}@{\hspace{5pt}}>{$}r<{$} >{\hspace{0pt}}c}
\hline\noalign{\smallskip}
\multicolumn{1}{l}{Experiment} & \multicolumn{1}{l}{Target} &
Events & \multicolumn{1}{r}{Determined from} & \multicolumn{1}{r}{$M_A,\,\GeV$} & {Ref.} \\
\noalign{\smallskip}\hline\noalign{\smallskip}
BNL 80       & Hydrogen  &        & \tdcs & \tnA{0.9^{+0.4}_{-0.3}}
                                  & \cite{Fanourakis:1980si} \\
BNL 88       & Liquid scint. & 2919 & \tdcs & \tnA{1.09 \pm 0.04}
                                  & \cite{Ahrens:1988rr} \\
\noalign{\smallskip}\hline\noalign{\smallskip}
FermiLab 84  & Neon      &    405 & \tdcs & \tnA{0.99\pm0.11}
                                  & \cite{Asratian:1984ir} \\
NuTeV 04     & Steel     &  15054 & \tcs  & \tnA{1.29 \pm 0.11}
                                  & \cite{Suwonjandee:04} \\
\noalign{\smallskip}\hline\noalign{\smallskip}
CERN GGM 77  & Freon     &    476 & \tlB & \tnB{0.69 \pm 0.44}{0.94 \pm 0.17}
                                  & \cite{Bonetti:1977cs} \\
CERN GGM 79 & Propane/Freon & 766 & \tlB & \tnB{0.84^{+0.08}_{-0.09}}{0.91 \pm 0.04}
                                  & \cite{Armenise:1979zg} \\
\noalign{\smallskip}\hline\noalign{\smallskip}
IHEP 85      & Aluminium    & 854 & d\sigma_{\nu+\bar{\nu}}/dQ^2 & \tnA{1.00 \pm 0.04}
                                  & \cite{Belikov:1983kg} \\
IHEP SKAT 88 & Freon        &  52 & \tlD & \tnA{0.72 \pm 0.23}
                                  & \cite{Grabosch:1986js} \\
IHEP SKAT 90 & Freon        &     & \tlC & \tnC{0.62\pm0.16}{0.79\pm0.11}{0.71\pm0.10}
                                  & \cite{Brunner:1989kw} \\
\noalign{\smallskip}\hline
\end{tabular}
\end{table*}

Let us start with 
a brief review of existing experimental data on (anti)neutrino
nucleon QEL scattering.

A compilation of available data on the cross-section measurement 
of the $\nu_{\mu}$ and $\bar\nu_{\mu}$ quasi-elastic
scattering off deuterons and other nuclei or composite targets 
(like freon, propane, liquid scintillator) 
as a function of the incoming neutrino energy has been made
(see Figures~\ref{fig:nomad_cc_nucl},~\ref{fig:nomad_cc_free} 
and~\ref{fig:nomad_ac_nucl}).
This study allowed to
conclude that
the QEL cross-section measured in different experiments can vary by 20-40\%. 

The existing data on (anti)neutrino QEL scattering come mostly from bubble chamber
(BC) experiments. In general, these data suffer from small statistics. Moreover,
results of several old
experiments~\cite{Orkin-Lecourtois:1967,Holder:1968,Budagov:1969bg} 
have large systematic uncertainties due to the poor knowledge of the incoming neutrino 
flux and of background contamination in the selected events.

The total QEL cross-section was recently measured in data collected by the NuTeV
collaboration~\cite{Suwonjandee:04}. The number of QEL events identified in their
analysis are comparable with the total world data obtained in previous experiments.
However, the results reported for the antineutrino case fall well outside the most
probable range of values known today and hence, seem to exhibit a systematic
shift. 

Another intriguing subject in the study of the neutrino quasi-elastic scattering is
the axial structure of the nucleon. We will skip here the details of the phenomenology
of the hadronic current involved in the matrix element of the process
(see Section~\ref{section:qel_mc} and 
 Ref.~\cite{Kuzmin:2007kr}).
But let us only remind the reader that for the region of low and intermediate 4-momentum transfer, 
$Q^2$, we can use a dipole parametrization for the axial form factor with only one adjustable
parameter, the so-called axial mass $M_A$.

The $M_A$ parameter describes the internal structure of the nucleon 
and should be the same both for neutrino and antineutrino experiments (if we assume
the isotopic invariance of strong interaction). Therefore, it is convenient to compare
experimental results in terms of the axial mass. 
There is, however, no theoretical basis for this form of the axial form factor.
The use of an inappropriate parametrization could therefore lead to values of 
$M_A$ that differ when extracted under different kinematical conditions.

There are two possible ways generally used to extract the $M_A$ parameter from
experimental data: 
\begin{enumerate}
\item from the total QEL (anti)neutrino nucleon cross-section (the axial form factor is
  responsible for about 50-60\% of the total QEL cross-section); 
\item from the fit of the $Q^2$ distribution of the identified neutrino QEL events.
\end{enumerate}

In principle, these two procedures should give self-consistent results. However,
the old bubble chamber experiments at ANL and CERN reported in general larger
values of $M_A$ based on the $Q^2$ fit than 
those obtained from the total cross-section
measurements. 

Results of the $M_A$ measurements based on the $Q^2$ fit have been 
recently published by the K2K~\cite{Gran:2006jn,Mariani:2008zz} and 
MiniBooNE~\cite{MiniBooNE:2007ru} collaborations.
They are about $15\%$ higher than the average of previous 
deuterium filled bubble chamber experiments. 
This disagreement is, however, just at about one sigma level
because of the large systematic errors associated with 
the K2K and MiniBooNE measurements.

Let us note that the extraction of $M_A$ from the $Q^2$ distribution fit is a more
delicate issue than the QEL total cross section measurement.

In general, there are at least three aspects which can affect noticeably the 
$M_A$ measurements:
\begin{enumerate}
\item
  The nuclear effects can distort the expected distributions of the measured
  kinematic variables (like the energy of the outgoing nucleon). The
  neutrino-nucleus interactions should be described by a theoretical model suitable 
  for the considered neutrino energy region. This is important both for MC modeling in
  present-day neutrino experiments and for a proper interpretation of the results
  obtained earlier (with few exceptions for the deuterium filled bubble chambers).
\item
  The correct determination of the background contamination from both deep
  inelastic scattering and single pion production in the selected events is
  important for experiments operating with intermediate and high energy neutrino
  beams. 
\item
  The QEL reconstruction efficiency as a function of $Q^2$ for two-track events is not expected to be a
  flat function. It should drop both at small $Q^2$ due to the loss of low energy
  protons and at large $Q^2$ due to the loss of low energy muons. Effects which influence the efficiency of
  the low momentum particle reconstruction should be carefully taken into account in
  the MC modeling of the detector response. 
\end{enumerate}

Table~\ref{tab:ma_numu} and~\ref{tab:ma_anumu} display 
the measured values of $M_A$ from neutrino and antineutrino experiments
(this compilation is also presented in graphical form
in Fig.~\ref{fig:ma_exp}).
Whenever possible we provide also the $M_A$ measured from the
total cross-section.

From the results described above one can conclude that the presently available
experimental data on the neutrino QEL cross-section allow for a very wide spread of
the axial mass values, roughly from $0.7$ to $1.3\;\GeV$. Therefore the reliability of a
theoretical fit to these data is questionable and the uncertainty attributed to such a fit
should go beyond the averaged experimental statistical accuracy.
Nevertheless, the formal averaging of $M_A$ values from several early experiments was done 
by the authors of~\cite{Bernard:2001rs}: $M_A = 1.026\pm 0.021\;\GeV$.
This result is also known as the axial mass world average value. According
to~\cite{Budd:2003wb,Budd:2004bp,Bodek:2007ym} 
an updated world average value from $\nu_\mu$-Deuterium and pion
electroproduction experiments is $M_A = 1.014\pm 0.014\;\GeV$.

\clearpage

\clearpage
\section{\label{section:nomad} The NOMAD detector}

\begin{figure}
\begin{center}
\mbox{\epsfig{file=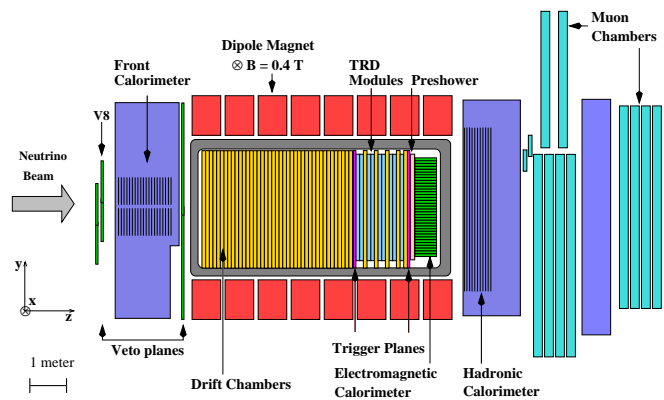,width=1.0\linewidth}}
\end{center}
\caption{\label{fig:nomad_detector}
A side-view of the NOMAD detector.
}
\end{figure}

The NOMAD detector~\cite{Altegoer:1997gv} consisted of an active target of 44 drift
chambers with a total fiducial mass of 2.7~tons, located in a 0.4~Tesla dipole
magnetic field as shown in Fig.~\ref{fig:nomad_detector}. The $X\times Y\times Z$
total volume of the drift chambers is about $300\times 300\times 400$ cm$^3$.

Drift chambers~\cite{Anfreville:2001zi}, made of low $Z$ material served the dual
role of a nearly isoscalar target\footnote{the NOMAD active target is nearly isoscalar
  ($n_n:n_p=47.56\%:52.43\%$) and consists mainly of Carbon; a detailed description of
  the drift chamber composition can be found in~\cite{Anfreville:2001zi}} 
for neutrino interactions and of tracking medium. The average density of the drift
chamber volume was 0.1 $\mbox{g}/\mbox{cm}^3$. These chambers provided an overall
efficiency for charged track reconstruction of better than 95\% and a momentum
resolution which can be approximated by the following formula
$\frac{\sigma_p}{p} \approx \frac{0.05}{\sqrt{L}} \oplus \frac{0.008p}{\sqrt{L^5}}$,
where the momentum $p$ is in GeV/c and the track length $L$ in m.
Reconstructed tracks were used to determine the event topology (the assignment of
tracks to vertices), to reconstruct the vertex position and the track parameters at
each vertex and, finally, to identify the vertex type (primary, secondary, etc.). 
A transition radiation detector (TRD)~\cite{Bassompierre1,Bassompierre2} placed at the end
of the active target was used for particle identification. 
Two scintillation counter trigger planes~\cite{Altegoer:1998qt} were used 
to select neutrino interactions in the NOMAD active target.
A lead-glass
electromagnetic  calorimeter~\cite{Autiero:1996sp,Autiero:1998ya} located downstream
of the tracking region provided an energy resolution of 
$3.2\%/\sqrt{E \mbox{[GeV]} } \oplus 1\%$ for electromagnetic showers and was crucial
to measure the total energy flow in neutrino interactions. In addition, an iron
absorber and a set of muon chambers located after the electromagnetic calorimeter was
used for muon identification, providing a muon detection efficiency of 97\% for
momenta greater than 5~GeV/c. 

The NOMAD neutrino beam consisted mainly of $\nu_\mu$'s with an about 7\% 
admixture of $\bar\nu_\mu$ and less than 1\% of $\nu_e$ and $\bar\nu_e$. 
More details on the beam composition can be found in~\cite{Astier:2003rj}.

The main goal of the NOMAD experiment was the search for neutrino
oscillations in a wide band neutrino beam from 
the CERN SPS~\cite{Astier:2001yj,Astier:2003gs}.
A very good quality of event reconstruction similar to that of bubble
chamber experiments and a large data sample collected during four
years of data taking (1995-1998) allow for detailed studies of
neutrino interactions.

\subsection{Reconstruction of QEL events in the NOMAD detector}

A detailed information about the construction and performance of the NOMAD
drift chambers as well as about the developed reconstruction algorithms 
is presented in~\cite{Anfreville:2001zi}. 
Let us briefly describe some features relevant to the current QEL analysis. 
The muon track is in general easily reconstructed.
However, when we study protons emitted in 
the $\nu_\mu$ QEL two-track candidates 
we deal with protons with momentum well below 1~GeV/c and
with emission angle above 60 degrees. For positive particles in the upward
hemisphere of the NOMAD detector such conditions mean that these particles
are almost immediately making a U-turn due to the magnetic field.
There were no special efforts invested into tuning the NOMAD
reconstruction program to reconstruct this particular configuration
(which is rather difficult due to the fact that these protons are in
the 1/$\beta^2$ region of ionization losses, 
traversing much larger amount of material,
crossing drift cells at very large angles where the spacial resolution 
of the drift chambers is considerably worse and where a large amount
of multiple hits is produced, etc.). 
Some of these effects are difficult to parametrize and to simulate 
at the level of the detector response in the MC simulation program.
Thus, the reconstruction efficiencies for this particular configuration 
of outgoing protons could be different for the simulated events and real data.

Let us stress, however, that for protons emitted downwards we observed 
a good agreement between data and MC.

In the current analysis it was important to disentangle the
reconstruction efficiency effects discussed above from the effects
induced by intranuclear cascade (which could change the proton kinematics
and thus introduce drastic changes in the final results due to the
efficiency mismatch between simulated and real data). 
In order to get rid of an interplay between these two effects it was crucial 
to choose the region in the detector with a stable reconstruction efficiency.
This could be achieved by selecting $\nu_\mu$ QEL events where protons 
are emitted in the lower hemisphere of the NOMAD detector. 
This approach allowed to find the best set of parameters for description of 
the intranuclear cascade.

The most upsteam drift chamber was used as an additional veto to remove 
through-going muons from neutrino interactions upstream of the NOMAD active 
target. This is crucial for the study of single track events.

\clearpage
\section{Monte Carlo simulation of neutrino interactions}
\label{section:nomad_mc}

Inclusive (anti) neutrino charged current (CC) and neutral current (NC) scattering can be
considered as a mixture of several processes described by significantly different
models. In our case, these are 
quasi-elastic scattering (QEL), single-pion
production (RES) and deep inelastic scattering (DIS). 
There is also a contribution from a coherent interaction 
of neutrino with a nucleus as a whole (COH).
Below we will describe in
details the simulation scheme used for each of these processes and discuss the
influence of the nuclear effects. 

An adequate MC description of neutrino interactions is important to calculate the
efficiency of the QEL selection. Moreover, it allows us to predict the level of
background, which cannot be suppressed completely by the QEL identification scheme
proposed in Section~\ref{section:QEL_selection}.

\subsection{Quasi-elastic neutrino scattering}
\label{section:qel_mc}

The standard representation of the weak hadronic current involved in the matrix 
elements of the processes $\nu_\mu n\to\mu^-p$ and $\overline{\nu}_\mu p\to\mu^+n$, 
is expressed in terms of 6 form-factors, which in general are assumed to be 
complex~\cite{Llewellyn_Smith:1971zm}. They formally describe the hadronic structure 
and cannot be calculated analytically within the framework of the electro-weak 
interaction theory.

We neglect the second-class current contributions associated with the scalar and
pseudo-tensor form-factors. This is equivalent to the requirement of time reversal 
invariance of the matrix element (hence all form-factors should be real functions 
of $Q^2$) and charge symmetry of the hadronic current (rotation about the second axis 
in the isotopic space).

The vector form-factors $F_V$ and $F_M$ are related through 
the isospin symmetry hypothesis 
to the electromagnetic 
ones, which we will consider to be well known. Instead of the simple dipole parametrization,
extensively used in previous experiments, we have chosen the Gari--Kr\"uempelmann (GK)
model~\cite{Gari:1992qw} extended and fine-tuned by Lomon~\cite{Lomon:2002jx}. 
Specifically we explore the ``GKex(05)'' set 
of parameters~\cite{Lomon:2006xb} which fits the modern and consistent older data well
and meets the requirements of dispersion relations and of QCD at low and high
4-momentum transfer~\cite{Gari:1992qw}. 

For the axial and pseudoscalar form factors we use the conventional
representations~\cite{Llewellyn_Smith:1971zm}:
\begin{equation}
F_A\left(Q^2\right)=F_A(0)\left(1+\frac{Q^2}{M^2_A}\right)^{-2}
\end{equation}
and
\begin{equation}
F_P\left(Q^2\right)=\frac{2m_N^2}{m^2_\pi+Q^2}F_A\left(Q^2\right),
\end{equation}
where $F_A(0)=g_A=-1.2695\pm 0.0029$ (measured in neutron
$\beta$-decay~\cite{Eidelman:2004wy}); $m_\pi$ and $m_N$ - pion and nucleon masses.

As discussed in Section~\ref{section:exp_data}, the currently available
experimental data on the axial mass $M_A$ allow for a wide spread. Thus, in our case,
it should be considered as one of the available parameters, which can be used to
adjust the MC simulation with the measured value of the total QEL cross section and
observed distributions of the kinematic variables (other parameters, related to the
modeling of the intranuclear cascade, will be described later).

Note that the expression for the pseudoscalar form factor $F_P$ is nothing better 
than a plausible parametrization inspired by the PCAC hypothesis and the assumption 
that the pion pole dominates at $Q^2 \lesssim m_\pi^2$~\cite{Llewellyn_Smith:1971zm}. 
However, its contribution enters into the cross sections multiplied
by a factor $(m_\mu/m_N)^2$. Hence, the importance of the related uncertainty is much
reduced.

\subsection{Single-pion production through intermediate baryon resonances}
\label{section:res_mc}

In order to describe the single-pion neutrino production through baryon resonances we
adopt an extended version of the Rein and Sehgal model
(RS)~\cite{Rein:1980wg,Rein:1987cb}, which seems to be one of the most widely trusted
phenomenological approaches for calculating the RES cross sections. The
generalization proposed in~\cite{Kuzmin:2003ji,Kuzmin:2004ya} takes into account the
final lepton mass and is based upon a covariant form of the charged leptonic current
with definite lepton helicity. In our MC simulation we use the same set of 18
interfering nucleon resonances with masses below $2~\GeV$ as in~\cite{Rein:1980wg} but 
with all relevant input parameters updated according to the current 
data~\cite{Eidelman:2004wy,k2k:2008ea}. Significant factors (normalization coefficients etc.),
estimated in Ref.~\cite{Rein:1980wg} numerically are recalculated by using the new
data and a more accurate integration algorithm.

The relativistic quark model of Feynman, Kislinger, and Ravndal~\cite{Feynman:1971wr},
adopted in the RS approach, unambiguously determines the structure of the transition
amplitudes involved into the calculation and the only unknown structures are the
vector and axial-vector transition form factors $G^{V,A}(Q^2)$.
In~\cite{Rein:1980wg} they are assumed to have the form
\begin{equation}\label{RS_Gva}
\frac{G^{V,A}(Q^2)}{G^{V,A}(0)} = 
\left(1+\frac{Q^2}{4m_N^2}\right)^{1/2-n}
\biggl(1+\frac{Q^2}{M_{V,A}^2}\biggr)^{-2}
\end{equation}
where the integer $n$ in the first (``ad hoc'') factor in Eq.~(\ref{RS_Gva})
is the number of oscillation quanta of the intermediate resonance.

The vector mass $M_V$ is taken to be $0.84~\GeV$, that is the same as in the usual 
dipole parametrization of the nucleon electromagnetic form-factors. The axial mass
(which was fixed at $0.95~\GeV$ in the original RS paper) is set to the standard
world averaged value $M_A=1.03~\GeV$. It is in good agreement with the results
obtained in the recent analysis of the data from the BNL 7-foot deuterium filled bubble
chamber~\cite{Furuno:2003ng} ($M_A=1.08\pm 0.07~\GeV$). Let us also note that the
available experimental data for the single-pion neutrino production (as in the case of QEL
scattering) does not permit a very definite conclusion about the value of the total RES 
cross section (and the corresponding axial mass value). The present uncertainties will 
be taken into account in the calculation of the systematic error of the current analysis.

To compensate for the difference between the $SU_6$ predicted value ($-5/3$)
and the experimental value for the nucleon axial-vector coupling $g_A$,
Rein and Sehgal introduced a renormalization factor $Z=0.75$.
In order to adjust the renormalization to the current world averaged value
$g_A=-1.2695$~\cite{Eidelman:2004wy} we have adopted $Z=0.762$.
The harmonic-oscillator constant $\Omega$, which accounts for the mass differences
between states with different number of excitation quanta is set to
its original value $\Omega = 1.05~\GeV^2$.

Another essential ingredient of the RS approach is the non-resonant background (NRB).
Its contribution is important in describing the existing data on the reactions
${\nu}_{\mu}n\to\mu^-n\pi^+$, ${\nu}_{\mu}n\to\mu^-p\pi^0$,
$\bar{\nu}_{\mu}p\to\mu^+p\pi^-$ and $\bar{\nu}_{\mu}p\to\mu^+n\pi^0$.
In our Monte Carlo, the NRB is taken to come from the DIS part of the
simulation. Therefore it has not been used in the RES part of our event generator.

\subsection{Deep inelastic scattering}
\label{section:dis_mc}

\begin{figure*}
\begin{center}
\mbox{\epsfig{file=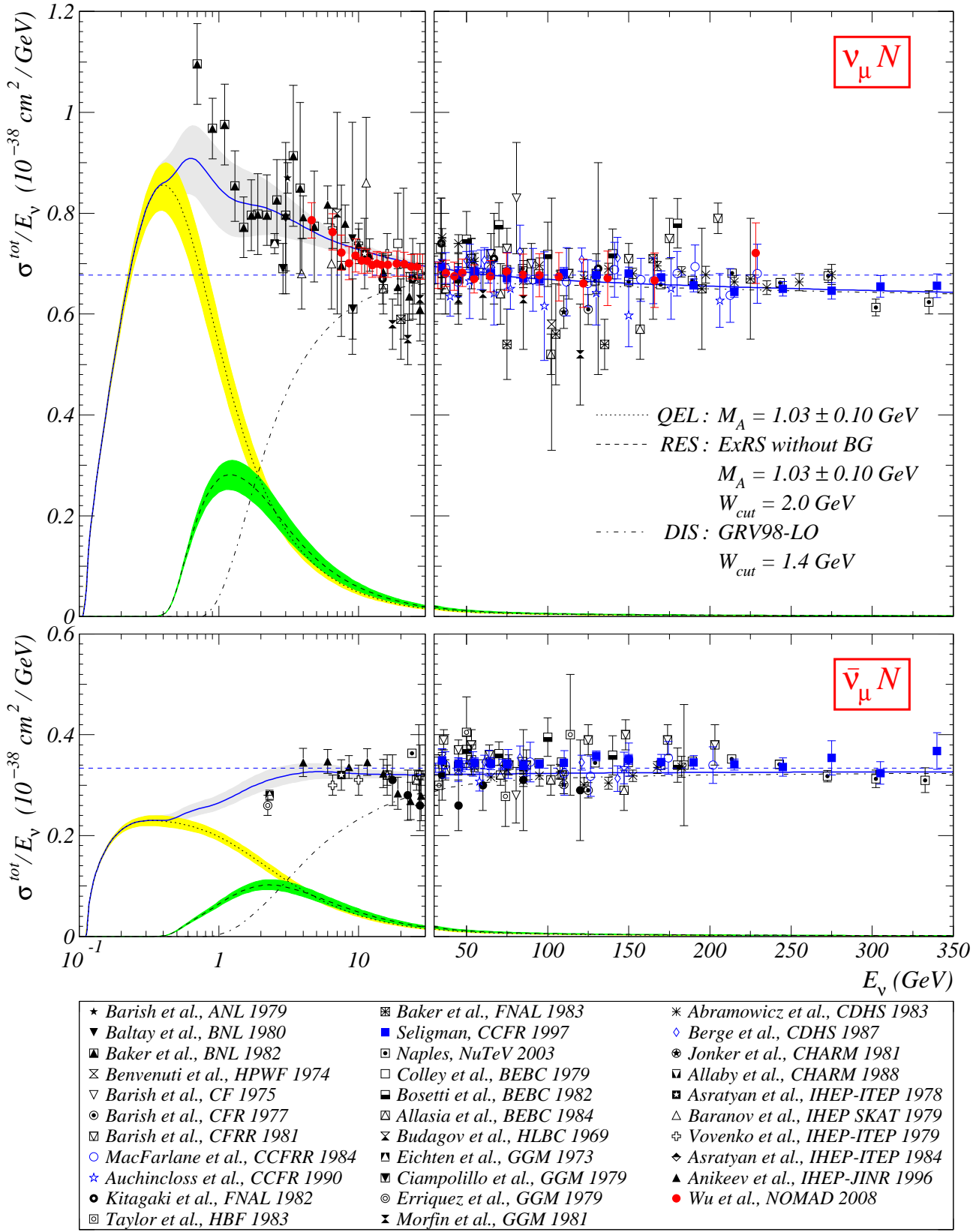,width=0.95\linewidth}}
\end{center}
\caption{\label{fig:cs_dis} 
Slopes of the total $\nu_{\mu}$ and $\overline{\nu}_{\mu}$ CC 
scattering cross-sections off an isoscalar nucleon 
(the compilation of experimental data is taken from~\cite{Kuzmin:2005bm}).
The curves and bands show the QEL, RES, and DIS contributions and their sums 
calculated with the parameters described in the legend of the top panel.
The averaged values over all energies 
$(0.677\pm0.014)\times10^{-38}~\mathrm{cm}^2/\mathrm{GeV}$ 
(for ${\nu}_{\mu}N$) and
$(0.334\pm0.008)\times10^{-38}~\mathrm{cm}^2/\mathrm{GeV}$ 
(for $\overline{\nu}_{\mu}N$) obtained by the
Particle Data Group~\cite{Eidelman:2004wy} are also shown for comparison (straight
lines). 
}
\end{figure*}

The MC simulation of the deep inelastic neutrino nucleon scattering is based on the
LEPTO~6.5.1 package~\cite{Ingelman:1996mq} with several
modifications~\cite{Levy:2004rk,Ellis:2002zv}. For hadronization we use the LUND string
fragmentation model, as incorporated into the JETSET~7.4 
program~\cite{Sjostrand:1995iq,Sjostrand:1985ys,Sjostrand:1986hx}.

Upon implementing the Monte Carlo for $\nu_{\mu}$($\bar{\nu}_{\mu}$) CC
scattering, kinematic boundaries between exclusive (RES) and inclusive (DIS) 
channels must be defined. To avoid double counting, the phase
space of the RES and DIS contributions should be separated by the conditions
$W<W_{cut}^{RES}$ and $W>W_{cut}^{DIS}$, where $W$ is the invariant mass of the final
hadronic system. 

The maximum possible value for $W_{cut}^{RES}$ is the upper limit of the RS model 
($2~\GeV$), while inelastic scattering can take place from the
one-pion 
production threshold 
(note, however, that this value is too small in principle since the structure functions 
used in the calculation of the DIS cross section cannot be extrapolated down 
to this value).

Unfortunately, there is no clear physical recipe to determine exact numerical values for
those cutoff parameters. The authors of GENIE MC code~\cite{Andreopoulos:2006cz} 
adopt the value $W_{cut}^{RES} \simeq W_{cut}^{DIS} \sim 1.7~\GeV$. 
A comprehensive analysis of available experimental data 
made in~\cite{Kuzmin:2005bm,Kuzmin:2006dt} suggests to decrease this cut to $\sim
1.5~\GeV$. 

In the present analysis we set $W_{cut}^{RES}=2~\GeV$ and $W_{cut}^{DIS}=1.4~\GeV$.
This choice allows for the non-resonant contribution to single pion production to be
accounted for by the DIS part of the Monte Carlo,
providing e.g. $N(\mu^-p\pi^0 \mbox{ in DIS})/N_{dis} \cdot \sigma_{dis} \approx \sigma (\mu^-p\pi^0 \mbox{ in NRB from RES})$, see previous subsection. 
Moreover, it is not at variance with experimental data as far as the total (anti)neutrino 
cross-section is concerned  (see Fig.~\ref{fig:cs_dis}).

\subsection{Coherent pion production}
\label{section:coherent_pion_production}

In the processes described above, neutrinos interact with individual target 
nucleons. However, pions can be produced in a coherent interaction 
of the neutrino with the whole nucleus, i.e. in the case of CC $\nu_\mu$ 
scattering $\nu_\mu\; \mathcal{N}\to \mu^-\pi^+ \mathcal{N}$, where $\mathcal{N}$ 
is the target nucleus.

The details of the MC simulation can be found in~\cite{Winton:1999zz}, 
which is devoted to the investigation of this process in the NOMAD experiment.
The flux averaged cross-section has been calculated following 
~\cite{Rein:1982pf,Rein:1986cd} and has been estimated at $0.733\times 10^{-38}\cm^2$ 
per nucleus. 
For a recent experimental result at low incoming neutrino energy 
see~\cite{Hiraide:2008eu}.
Taking into account that the average mass number of the NOMAD 
target is 12.9, and using the number of recorded DIS events (see 
section~\ref{section:dis_norm}) one finds that the expected number of coherent
pion production events is $\sim 2700$. Nevertheless, the probability for events of this  
type to be identified as QEL is $\sim 2\%$ because of the small pion emission angle,  
so that the expected contamination of the selected QEL sample is lower than $0.4\%$.

\hfill

\subsection{Nuclear effects}
\label{section:nuclear_effects}

\begin{figure}
\begin{center}
\begin{tabular}{c}
\mbox{\epsfig{file=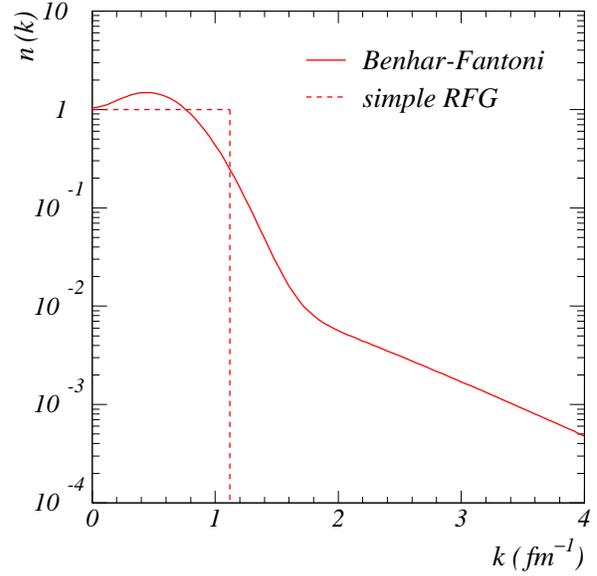,width=0.90\linewidth}}
\end{tabular}
\end{center}
\caption{\label{fig:BF}
Benhar-Fantoni parametrization~\cite{Benhar:1994hw} for the momentum distribution
of the target nucleons (solid line), normalized to the Fermi distribution with
zero temperature and Fermi momentum $P_F = 221\,\MeV/c$ (simple RFG, dashed line).
}
\end{figure}

For typical NOMAD neutrino energies, we can assume that the incident neutrino
interacts with one nucleon only inside the target nucleus, while the remaining nucleons
are spectators (Impulse Approximation). In this case, one can describe the neutrino
nucleus scattering by folding the usual expressions for the free neutrino nucleon
cross sections with a Fermi gas distribution.

In the relativistic Fermi gas model, the nucleus is considered as an infinite
system of non-interacting nucleons. The phenomena related to the nuclear surface and
to the interaction between nucleons can be taken into account by using a more realistic 
effective momentum distribution for the target nucleons. In the NOMAD event generator
we used the Benhar-Fantoni parametrization~\cite{Benhar:1994hw}, see Fig.~\ref{fig:BF}. 

The QEL simulation is based on the Smith-Moniz approach~\cite{Smith:1972xh}. The
momentum of the recoil nucleus and the nucleon binding energy are included in the
conservation laws which determine the event kinematics. The only final state
interaction (FSI) effect which is taken into account at this stage is the Pauli
exclusion principle. The explicit form of the QEL differential cross section
used in the MC code can be found in~\cite{Kuzmin:2007kr}. 

MC implementation of the Fermi gas model in the case of single pion production is more
straightforward. First, we generate the momentum of the target nucleon and
make a Lorentz boost to its rest frame where the RES event can be simulated 
according to the extended RS model described in subsection~\ref{section:res_mc}.
The effect of Pauli blocking on the outgoing nucleon is taken into account as it is 
in the QEL MC.

In the case of the DIS neutrino scattering there are several specific nuclear effects 
(such as nuclear shadowing, pion excess and off-shell corrections to bound nucleon
structure functions). They are described in the theoretical framework proposed
in~\cite{Kulagin:2004ie}.

Simulating the re-interactions between particles produced at the primary 
neutrino collision off the target nucleon with the residual nucleus is an important
ingredient of the MC event generator. To include this effect, commonly called final
state interactions, we use the DPMJET package~\cite{Battistoni:1998hh}. 

The intranuclear re-interaction of the particles generated by the QEL, RES or DIS 
event generators can be described and simulated by the Formation Zone Intranuclear 
Cascade model~\cite{Ranft:1988kc,Ferrari:1995cq} implemented in DPMJET. Secondaries from 
the first collision are followed along straight trajectories and may induce in turn
intranuclear cascade processes if they reach the end of their ``formation zone'' inside
the target; otherwise they leave the nucleus without interacting. 

There are two important parameters in DPMJET. The first one, called the formation
time $\tau_0$, controls the development of the intranuclear cascade. With increasing
$\tau_0$, the number of cascade generations and the number of low-energy particles 
will be reduced. Its default value is $\tau_0=2.0$.
After some tuning described below we adopted the value $\tau_0=1.0$
in our simulation of QEL, RES and DIS events.

Inside DPMJET, the momenta of the spectator nucleons are sampled from the zero
temperature Fermi-distribu\-tion. However, the nuclear surface effects and the
interaction between nucleons result in a reduction of the Fermi momentum, see
Fig.~\ref{fig:BF}. It can be accounted for by introducing a correction factor
$\alpha^F_{mod}$ (default value 0.6). Moreover, $\alpha^F_{mod}$ provides the
possibility of some modification of the momentum distribution for the emitted
low-energy nucleons. 

At the end of the intranuclear cascade, the residual nucleus is supposed to go through 
some de-excitation mechanisms. It can be disrupted into two or more fragments, 
emit photons, nucleons or light particles (like $d$, $\alpha$, $^3\rm{H}$,
$^3\rm{He}$). We can easily neglect this contribution, since the typical energy of
those particles is below the observation threshold of the NOMAD detector.

In our analysis, special attention will be devoted to the dependence of the
obtained results on the intranuclear cascade parameters. As a cross-check, we 
compare our MC simulation for the QEL process with the predictions of the NUANCE event
generator~\cite{Casper:2002sd}, which is currently used in a large number of neutrino
experiments 
and which contains a different approach to the modeling of FSI effects.

\subsection{Expected signal/background ratio in the $\nu_{\mu}(\bar{\nu}_\mu)~CC$ sample}
\label{section:acs}

\begin{figure}
\begin{center}
\begin{tabular}{c}
\mbox{\epsfig{file=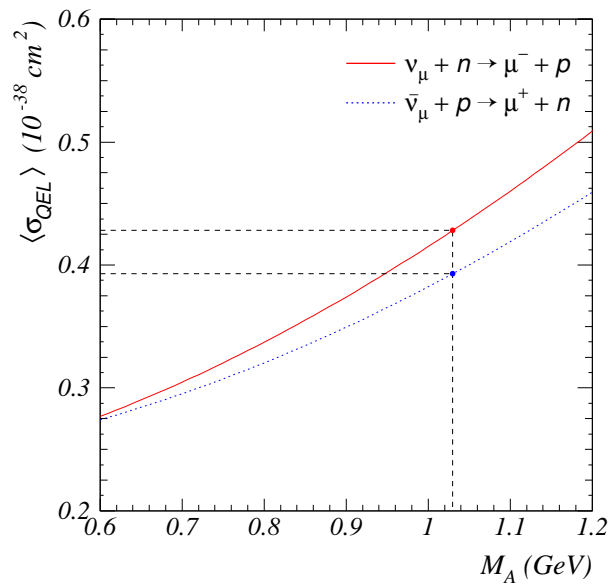,width=0.90\linewidth}}
\end{tabular}
\end{center}
\caption{\label{fig:acs_qel}
Flux averaged cross-section of QEL (anti)neutrino scattering for NOMAD
$\nu_\mu(\bar{\nu}_\mu)$ beam as a function of the axial mass $M_A$.
}
\end{figure}

In this subsection we estimate the number of signal quasi-elastic events in the initial
$\nu_{\mu}(\bar{\nu}_\mu)$ CC sample.

The contribution of each process to the total set of events is proportional to its
flux averaged cross-section:
\begin{equation}
\langle \sigma \rangle = 
\int \sigma(E_{\nu}) \Phi(E_{\nu}) dE_{\nu}\bigg / \int \Phi(E_{\nu}) dE_{\nu}
\label{eq:acs}
\end{equation}
where 
\begin{equation*}
\sigma(E_{\nu}) = n_n \sigma_{\nu n}(E_{\nu}) + n_p \sigma_{\nu p}(E_{\nu})
\end{equation*}
is the theoretical prediction for the cross-section of the process at stake, 
$\Phi(E_{\nu})$ denotes the NOMAD (anti)neutrino energy spectrum\footnote{the
  procedure used for the calculation of the flux and composition of the CERN SPS
  neutrino beam is described in~\cite{Astier:2003rj}}; 
$n_n$($n_p$) is the relative fraction of neutrons(protons) in the NOMAD target (see
Section~\ref{section:nomad}). 

The QEL cross-section was calculated in the framework of the Smith and Moniz
model~\cite{Smith:1972xh} for Carbon with binding energy $E_b = 25.6\;\MeV$ and Fermi
momentum $P_F = 221\;\MeV/c$. As noted above, the final result depends strongly on the
axial mass $M_A$ (see Fig.~\ref{fig:acs_qel}). 

To estimate the RES contribution, we fold the extended RS model~\cite{Kuzmin:2004ya}
for a free nucleon with the Pauli factor from~\cite{Paschos:2003qr}. The computation 
of $\sigma_{dis}(E_\nu)$ has been done with the GRV98-LO PDF model as indicated 
in~\cite{Kuzmin:2005bm}. The cutoff parameters $W_{cut}^{RES}$ and
$W_{cut}^{DIS}$ are the same as for the MC simulation.

Table~\ref{table:cs_all} contains our results for the reduced fiducial volume of the
NOMAD detector: $|X,Y|\leqslant 100~\cm$; the average $\nu_\mu~(\bar{\nu}_\mu)$ energy
was $25.9~(17.6)~\GeV$. 

Combining all these, the expected fraction of quasi-elastic events in the initial
$\nu_{\mu}(\bar{\nu}_{\mu})$ CC sample before any special selection is about
2.4\%(6.9\%) or $\sim 20300$($\sim 1360$) events. 

\begin{table}[t]
\caption{\label{table:cs_all}
Flux averaged cross-sections of the QEL, RES, DIS CC and NC processes per one nucleon of the NOMAD target.
Neutrino beam spectrum corresponds to the $|X,Y|\leqslant 100\, \cm$ fiducial area. The unit used for the cross-section is $10^{-38}\, \cm^2$. \hfill
}
\begin{tabular}{l >{\hspace{10pt}}r >{\hspace{10pt}}r}
\hline\noalign{\smallskip}
Process type & $\nu_\mu$     & $\bar{\nu}_\mu$ \\
\noalign{\smallskip}\hline\noalign{\smallskip}
QEL      &     0.428     &           0.393 \\
RES      &     0.576     &           0.432 \\
DIS CC   &    16.643     &           4.876 \\  
\noalign{\smallskip}\hline\noalign{\smallskip}
DIS NC   &     5.335     &                 \\
\noalign{\smallskip}\hline
\end{tabular}
\end{table}

\section{\label{section:QEL_selection} Events selection}

In this section we describe particular features of reconstruction and 
identification of $\nu_\mu$ and $\bar \nu_\mu$ QEL events.

\begin{figure*}
\begin{center}
\begin{tabular}{cc}
\mbox{\epsfig{file=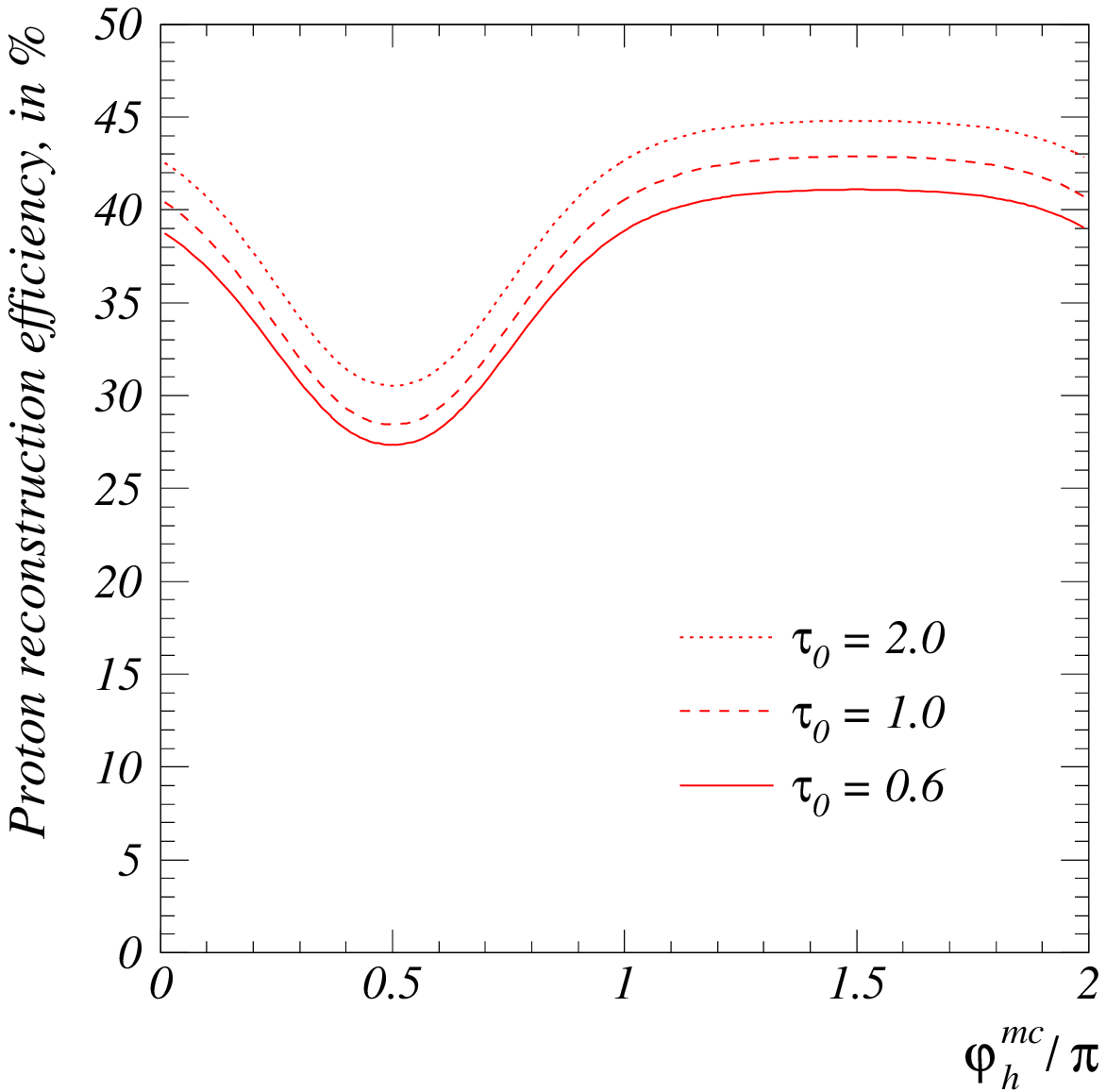,width=0.45\linewidth}} &
\mbox{\epsfig{file=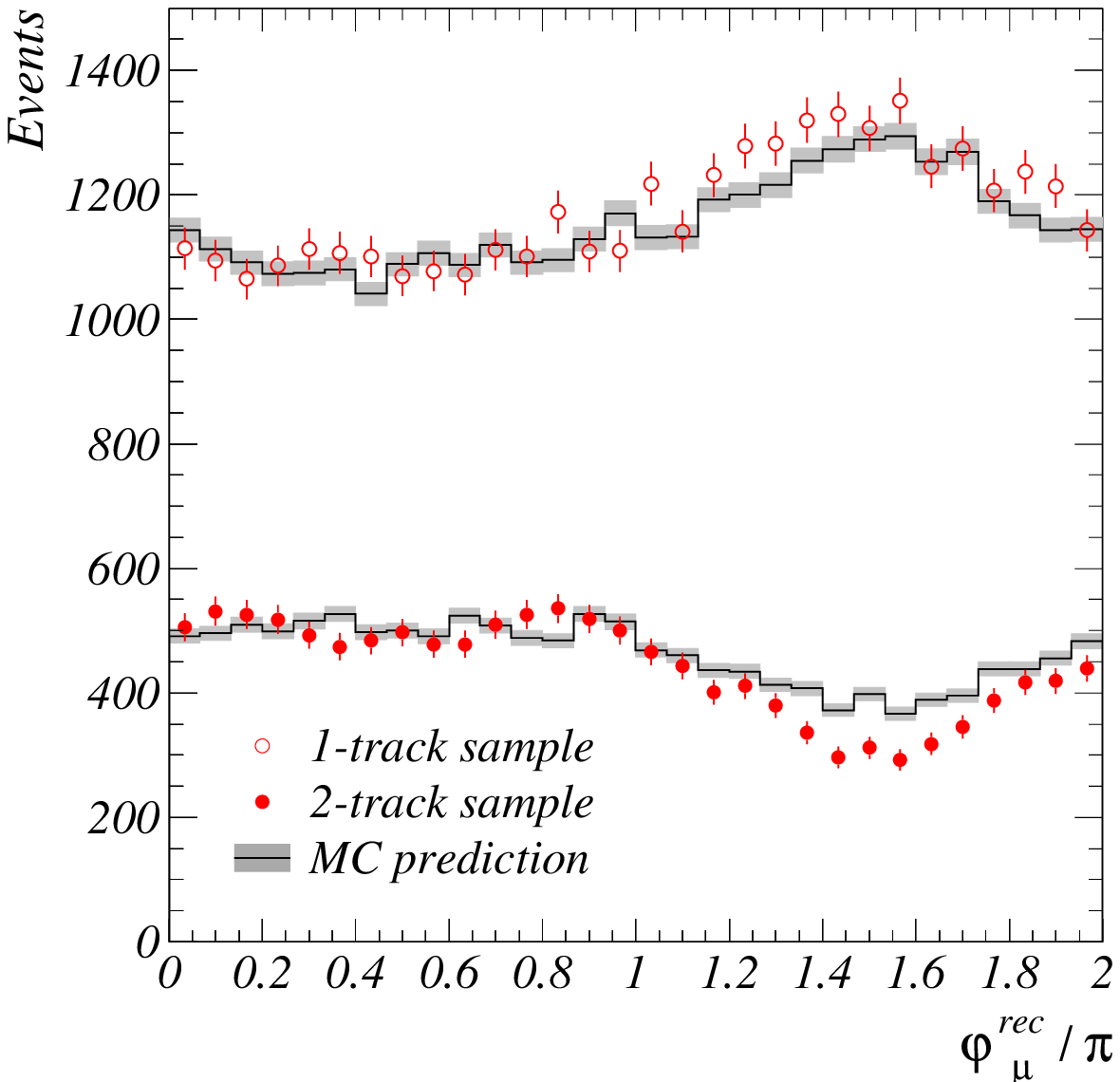,width=0.45\linewidth}}
\end{tabular}
\end{center}
\caption{\label{fig:eff_phi_pr}
The reconstruction efficiency of proton track as a function of its 
azimuth $\varphi_h$ for $\nu_\mu$ QEL scattering; the curves are smoothed 
MC predictions obtained for different values of the formation time $\tau_0$ (left).
The comparison of the muon azimuth $\varphi_\mu$ distributions 
in data and MC for 1-track and 2-track samples (right).
}
\end{figure*}

\subsection{$\nu_\mu n \to \mu^- p$ selection} 

For a $\nu_\mu n \to \mu^- p$ event one can expect two tracks originating
from the reconstructed primary vertex\footnote{all charged tracks originating within
  a $5~\cm$ box around the reconstructed primary vertex are forced to be included
  into it; we have also tried to vary this parameter by enlarging the size of the 
  box to $10~\cm$ and found that the final results are rather stable (within $0.3\%$
  for the measured QEL cross-section)}:
one of them should be identified as a muon, while the second track is assumed to be
a proton. Later we shall refer to events with such a topology as 2-track (two track)
events\footnote{in this analysis we do not take into account clusters in the
  electromagnetic calorimeter, which can be associated with neutral particles,
  originating from the primary vertex}. 

Sometimes the proton track cannot be reconstructed, 
e.g. if its momentum is below the
detector registration threshold. In this case, we deal with only one muon track
and we call such an event a 1-track (single track) event. 

The expected ratio between 1-track and 2-track events
for the pure standard QEL MC sample is $54.3\%:45.7\%$.

There are three possible reasons for the reconstruction of the proton track in a
QEL event to fail: 
\begin{itemize}
\item the proton, which was born in the neutrino interaction with the 
  target nucleon, has too low a momentum or too large an emission angle
  (this depends on the parameters of the model used to describe the neutrino-nucleon 
  interaction, in particular, on the value of the axial mass);
\item the proton from the primary neutrino interaction was involved in an 
  intranuclear cascade and lost part of its energy (this is controlled by the DPMJET
  parameters, mainly by the formation time $\tau_0$);
\item the detector magnetic field deviates positively charged particles upwards;
  therefore, if a slow proton is emitted at an azimuth $\varphi_h\sim \pi/2$, its 
  trajectory is almost parallel to the drift chamber planes and its track
  reconstruction efficiency
  (which depends on the number of hits associated 
  with the track) is significantly lower than in the case of a 
  proton emitted downwards at $\varphi_h\sim 3\pi/2$.
\end{itemize}

\begin{figure*}
\begin{center}
\begin{tabular}{cc}
\mbox{\epsfig{file=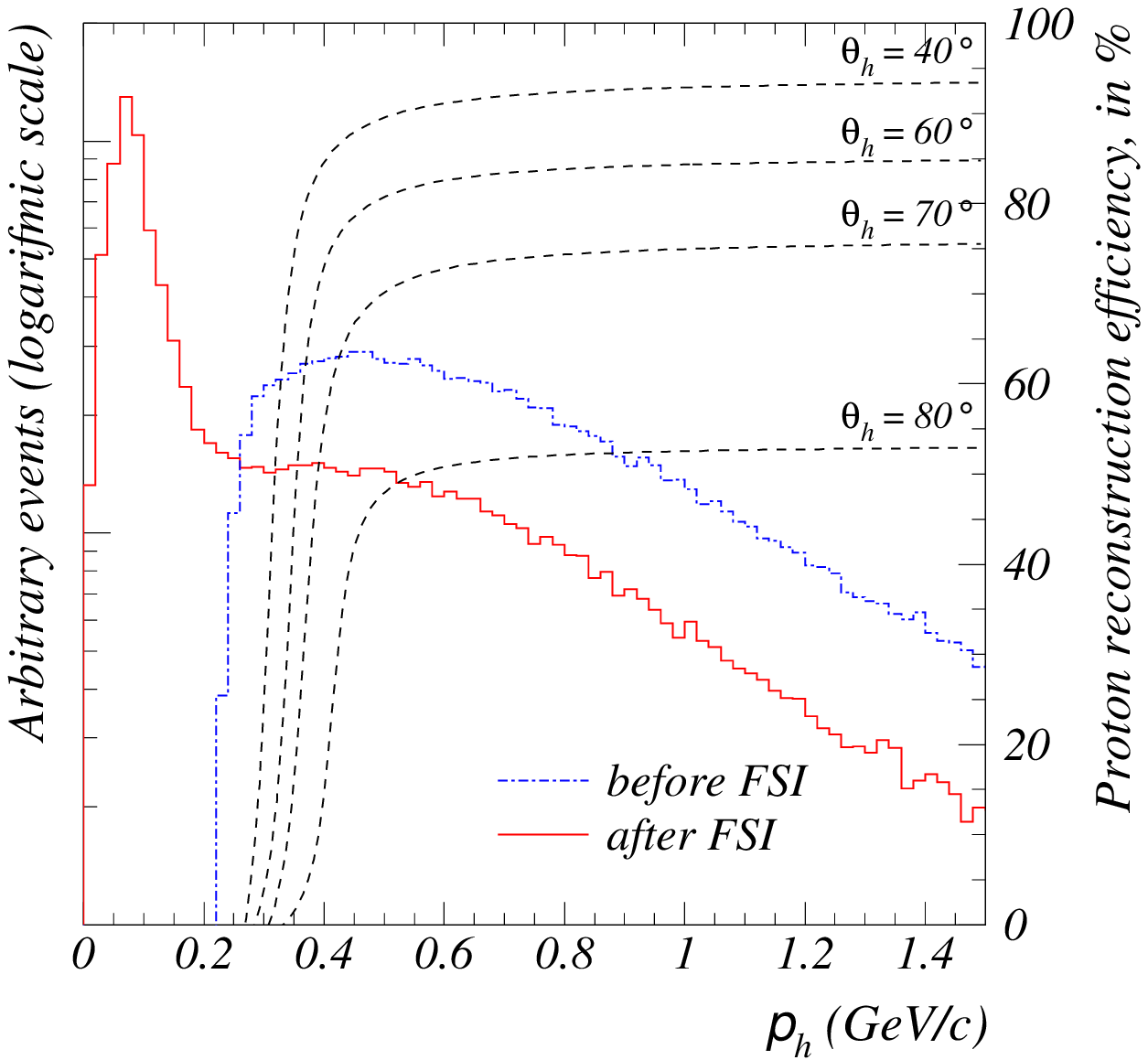,width=0.45\linewidth}} &
\mbox{\epsfig{file=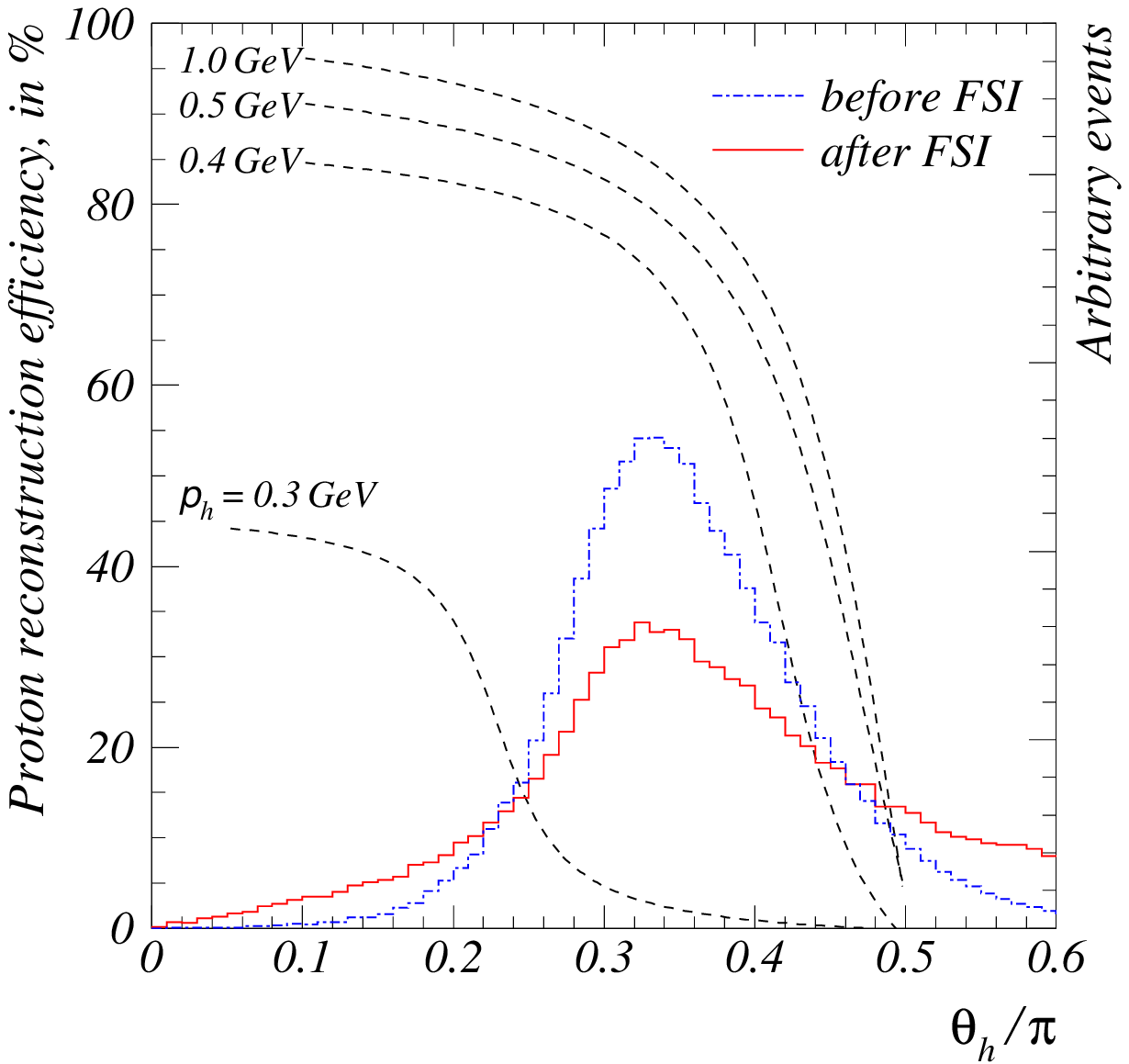,width=0.45\linewidth}}
\end{tabular}
\end{center}
\caption{\label{fig:eff_pr}
Distribution of the leading proton momentum (left) and emission angle (right) 
before (dash-dotted line) and after (solid line)  FSI simulation. Dashed 
lines show the proton reconstruction efficiency as function of the proton momentum 
and emission angle (for $\pi<\varphi_h<2\pi$).
}
\end{figure*}

In Fig.~\ref{fig:eff_phi_pr} (left) we illustrate these last two effects: 
the magnetic field is the cause of the asymmetry in the azimutal distribution
of the reconstructed protons, while variyng the formation time parameter $\tau_0$ 
affects the expected number of tracks uniformly. 

In Fig.~\ref{fig:eff_pr} we display an example of distributions of the leading proton
momentum $p_h$ and emission angle $\theta_h$ before and after FSI for the QEL 
neutrino scattering. The proton reconstruction probabilities are also shown 
as functions of $p_h$ and $\theta_h$: one can observe a fast decrease at low 
proton momenta (below $300~\MeV/c$) and large emission angles (larger than $72^o$). 
So, FSI tends to increase the fraction of events in kinematic domains with low
proton reconstruction efficiency and therefore to change the expected fraction of events
with a given topology in the identified QEL sample.

Using 2-track events only for the analysis may seem very attractive, since
we could significantly reduce the background contamination with the help of additional
kinematic variables (details can be found below). However, the results thus obtained
might still have large systematic uncertainties coming from insufficient understanding 
of nuclear effects.

The QEL events which are not reconstructed as 2-track events will populate mainly 
the 1-track sample. But $\sigma_{qel}$ extracted from this sample will suffer
from the same source of uncertainty. However, the measurement of the QEL cross-section 
simultaneously from both samples is expected to have only little dependence on
the uncertainties in the modeling of FSI effects and this is indeed what is found in the data 
(see Section~\ref{section:results}).

\begin{table*}
\caption{\label{tab:qel_identification} 
  Number of data $N_{data}$ and renormalized MC $N_{mc}$ events in $\nu_\mu$ and $\bar\nu_\mu$ QEL samples; 
  expected selection efficiency, purity and background contaminations (BG)
  for different stages of the analysis.
  \hfill}
\begin{tabular}{l>{\hspace{-10pt}}rrrrrrr}
\hline\noalign{\smallskip}
 & QEL eff. (\%) & QEL purity (\%) & RES BG (\%) & DIS BG (\%) & 
oth. BG (\%) & $N_{data}$ & $N_{mc} $\\
\noalign{\smallskip}\hline\noalign{\smallskip}
$\nu_\mu$ 1-track before $\theta_h$ cut       & 23.7 & 29.0 & 18.3 & 52.3 &    0.4 & 16508 & 16633.7 \\
$\nu_\mu$ 1-track after $\theta_h$ cut        & 21.3 & 41.7 & 23.2 & 34.5 &    0.6 & 10358 & 10358.0 \\
\noalign{\smallskip}\hline\noalign{\smallskip}
$\nu_\mu$ 2-track before $\mathcal{L}$ cut    & 17.6 & 47.2 & 17.3 & 35.2 &    0.3 &  7575 & 7609.0 \\
$\nu_\mu$ 2-track after $\mathcal{L}$ cut     & 13.3 & 73.9 & 10.2 & 15.8 & $<0.1$ &  3663 & 3663.0 \\
\noalign{\smallskip}\hline\noalign{\smallskip}
$\nu_\mu$ combined before cuts                & 41.3 & 34.7 & 18.0 & 47.0 &    0.3 & 24083 & 24242.7 \\
$\nu_\mu$ combined after cuts                 & 34.6 & 50.0 & 19.8 & 29.7 &    0.5 & 14021 & 14021.0 \\
\noalign{\smallskip}\hline\noalign{\smallskip}
$\bar{\nu}_\mu$ 1-track before $\theta_h$ cut & 81.8 & 29.8 & 22.8 & 45.8 &    1.6 &  3585 & 3555.8 \\
$\bar{\nu}_\mu$ 1-track after $\theta_h$ cut  & 64.4 & 36.6 & 28.5 & 33.6 &    1.3 &  2237 & 2237.0 \\
\noalign{\smallskip}\hline
\end{tabular}
\end{table*}


Therefore, the strategy of our analysis (selection criteria) in the case 
of $\nu_\mu n\to \mu^- p$ can be outlined as follows:
\begin{itemize}
\item {\em Fiducial volume cut.} 
  The reconstructed primary vertex should be within the restricted
  \footnote{we use a more stringent cut $Z>50~\cm$ for the data collected during 97 and 
    98, when the first drift chamber module was substituted by the NOMAD STAR detector}
  fiducial volume (FV):
  \begin{equation}
    |X,Y|\leqslant 100~\cm,\quad 25\leqslant Z \leqslant 395~\cm
    \label{eq:fv}
  \end{equation}
\item {\em Identified muon.} 
  We require the presence of a reconstructed and identified negatively
  or positively charged muon for the neutrino and antineutrino analyses respectively.
  In order to avoid possible problems with detector reconstruction inefficiencies, 
  we require $0<\varphi_\mu<\pi$, where $\varphi_\mu$ is the muon azimuthal angle (so, the proton
  track should lie in the bottom hemisphere), 
  see Fig.~\ref{fig:eff_phi_pr} (right).

This choice is 
validated
by our final errors being dominated by
systematics as we will be shown below.

\item {\em Event topology and reconstructed kinematic variables.} 
  We assign the events to the 1-track and 2-track subsamples and calculate $E_\nu$ 
  and $Q^2$.
  \begin{itemize}
    \renewcommand{\labelitemii}{\textbullet}
  \item {\em Single track sample} 
    (only one charged lepton is reconstructed and identified).
    To avoid contamination from the through-going muons we extrapolate 
    the muon track to the first drift chamber and require the absence of 
    veto chamber hits in the vicinity of the intersection point. The efficiency 
    of this quality cut was controlled by visual scanning of the reconstructed
    1-track events in the experimental data and was found to be satisfactory.
    Another quality cut was used to suppress a possible contribution from inverse 
    muon decay events: we require the muon transverse momentum to be greater
    than $0.2~\GeV/c$ (see Section~\ref{section:qel_cs} for more details).
    
    The kinematic variables are reconstructed under the assumption that the target nucleon 
    is at rest. For the 1-track events, the muon momentum and direction are the sole measurements 
    and we have to use the conservation laws (assuming QEL) to compute other kinematic quantities:
    \begin{align}
      & E_{\nu} = \dfrac{ME_\mu - m_\mu^2/2}{M-E_\mu+p_\mu\cos\theta_\mu} 
                                                         \nonumber \\
      & Q^2 = 2M(E_\nu-E_\mu)                                           \nonumber \\
      & p_h = ((E_\nu-p_\mu \cos\theta_\mu)^2
              + p_\mu^2 \sin^2\theta_\mu)^{1/2}                         \nonumber \\
      & \cos\theta_h = (E_\nu-p_\mu \cos\theta_\mu)/p_h,                
      \label{eq:fake_theta_pr}
    \end{align}
    where $p_\mu$, $\theta_\mu$ ($p_h$, $\theta_h$) are the momentum and emission 
    angle of the outgoing muon (nucleon), see Fig.~\ref{fig:like_var_cs}. 
    We note that for the neutrino energies relevant for this analysis 
    (above 3~GeV) there is no difference between the calculations based on 
    the approximated formulae above and the precise one, which takes into 
    account the binding energy (see e.g. Eq.~(4) in~\cite{MiniBooNE:2007ru}).
    With the help of the MC simulation we estimate 
    the resolution of the reconstructed $E_{\nu}$ and $Q^2$ as $3.6\%$ and $7.8\%$ 
    respectively.
    
  \item {\em Two track sample}
    (both the negative muon and the positively charged track are reconstructed).
    For a reliable reconstruction, we require that the number of hits 
    associated with the positively charged track should be greater than 7 
    and its momentum $p_h> 300~\MeV/c$. Otherwise such an event is downgraded 
    to the 1-track sample.
    
    For 2-track events, we use both the muon and the proton reconstructed 
    momenta to estimate $E_\nu$ and $Q^2$:
    \begin{align}
      & E_{\nu} = p_\mu \cos\theta_\mu + p_h \cos\theta_h                 \nonumber \\
      & Q^2 = 2 E_{\nu}(E_\mu-p_\mu \cos\theta_\mu)-m_\mu^2               \nonumber
    \end{align}
    The expected resolutions for $E_{\nu}$ and $Q^2$ are $3.6\%$ and $7.1\%$.
    
  \end{itemize}

\begin{figure}
\begin{center}
\begin{tabular}{c}
\mbox{\psfig{file=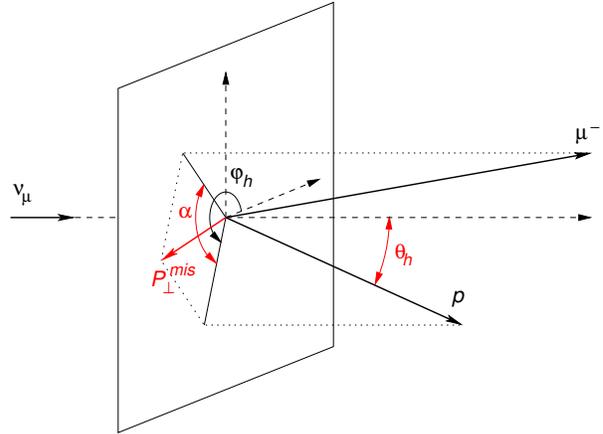,width=0.9\linewidth}}
\end{tabular}
\end{center}
\caption{\label{fig:like_var_cs}
Likelihood variables: missing transverse momentum $P_\perp^{mis}$, 
proton emission angle $\theta_h$, angle $\alpha$ between
the transverse components of the charged tracks.
}
\end{figure}

\begin{figure*}
\begin{center}
\begin{tabular}{cc}
\mbox{\epsfig{file=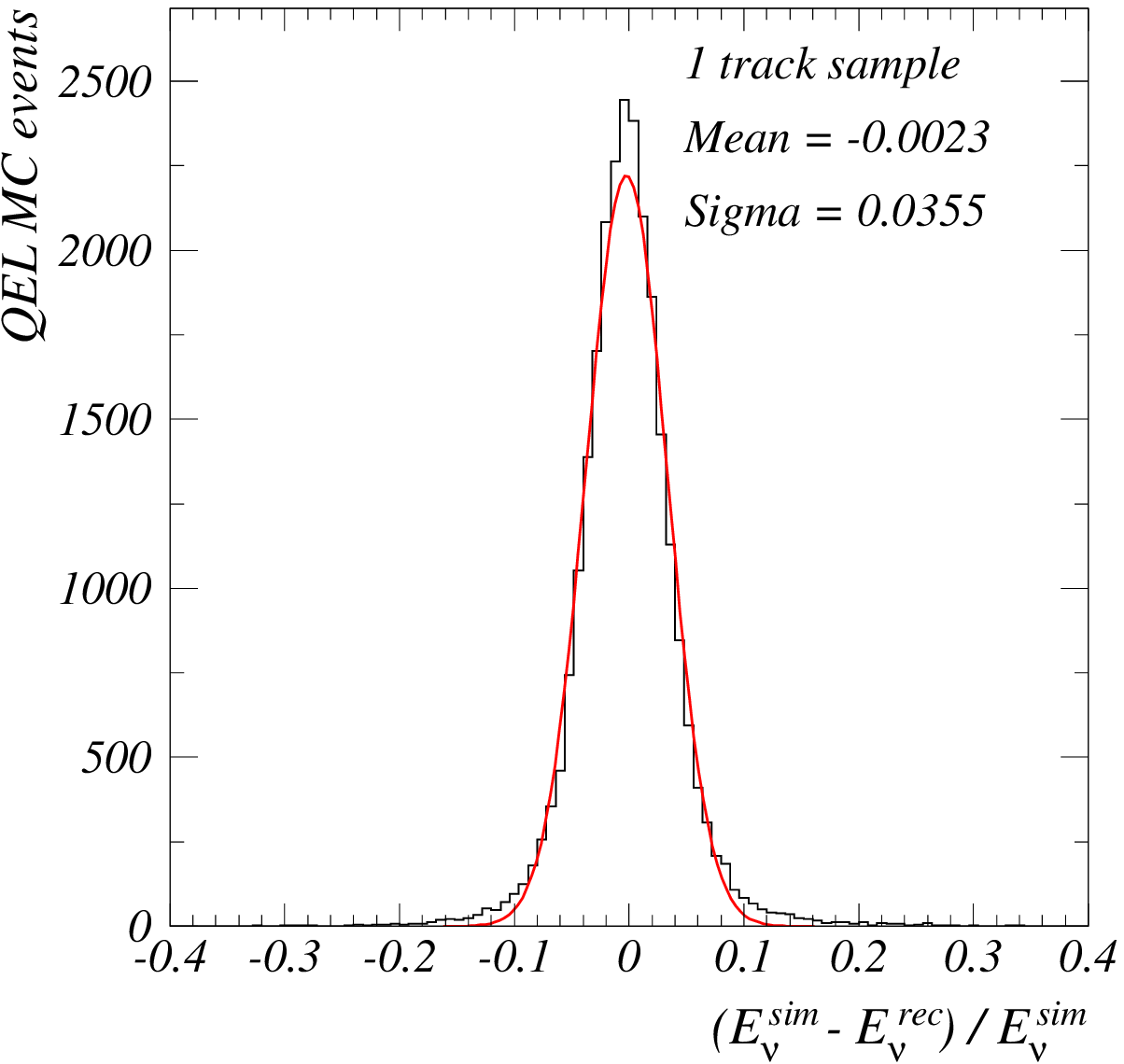,width=0.45\linewidth}} &
\mbox{\epsfig{file=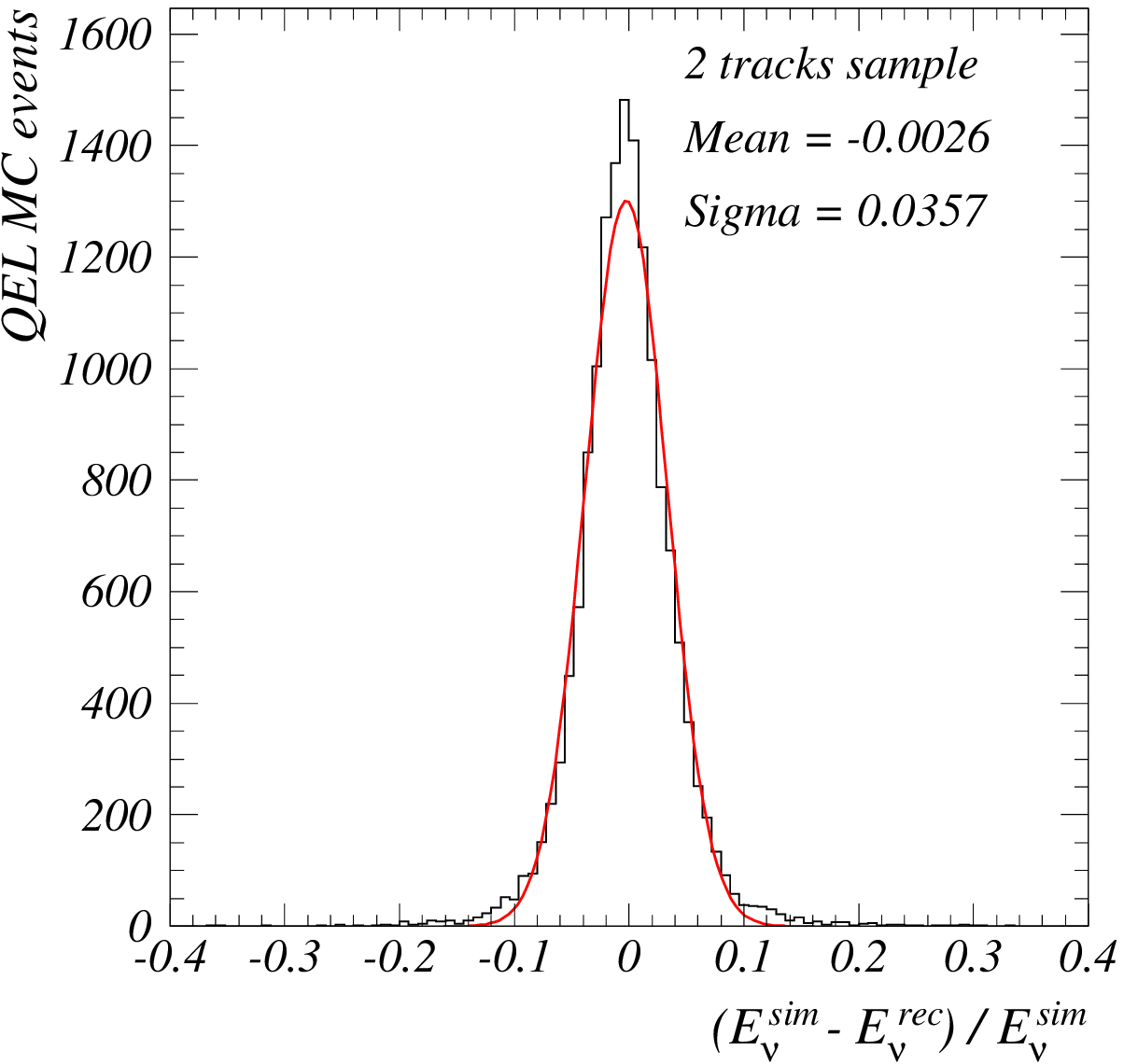,width=0.45\linewidth}}
\end{tabular}
\end{center}
\caption{\label{fig:enu_resolution}
The quality of the neutrino energy $E_{\nu}$ reconstruction
for 1- and 2-track samples.
}
\end{figure*}  

The quality of the neutrino energy $E_{\nu}$ reconstruction
for 1- and 2-track samples is illustrated in Fig.~\ref{fig:enu_resolution}.
It was checked that for the 2-track sample the derived cross-sections
are consistent within errors for both methods of $E_{\nu}$ calculation.

\item {\em Background suppression.}
  The contamination from RES and DIS processes can be suppressed by using the 
  difference between kinematical distributions in the QEL and background events
  as well as by the identification of the reconstructed positively charged 
  track as a proton (for the 2-track sample only). Therefore we apply:
  \begin{itemize}
    \renewcommand{\labelitemii}{\textbullet}
  \item {\em Identification of the positively charged track.}

    Momentum-range method~\cite{memo_98-023} can be reliably applied for low energy 
    protons since their tracks are shorter compared to that of $\pi^+$ (the main 
    background for proton identification) due to larger ionization 
    losses. In our case, this method can be applied to about $17\%$ of the events
    \footnote{We also undertook an attempt to identify positively charged particles 
      using the TRD information. A special algorithm 
      \cite{memo_97-028,memo_98-011}
      can be potentially used for discrimination between two particle-ID 
      hypotheses ($p/\pi$ in our case).  However, a low momentum ($\sim 0.9\, \GeV$) 
      of the particle and a rather large emission angle ($\gtrsim 45^{\circ}$) result 
      in that either the particle does not reach the TRD or the number 
      of residual TRD hits is not large enough for the identification. Therefore, the TRD 
      algorithm could be applied only to a limited fraction of events ($\sim 6\%$) 
      and cannot play any significant role in our analysis.
    }.

  \item {\em Kinematical criteria.}

    In the case of the 2-track sample, we can use additional kinematic 
    variables to suppress background contamination. We build the likelihood ratio 
    \begin{equation}
      \mathcal{L} = \ln \frac{P(\vec{\ell}\;|QEL)}{P(\vec{\ell}\;|BG)},
      \label{eq:like}
    \end{equation}
    using 3-dimensional correlations between the following kinematic variables 
    (see Fig.~\ref{fig:like_var_cs}):
    the missing transverse momentum $P_\perp^{mis}$,
    the proton emission angle $\theta_h$,
    and the angle $\alpha$ between the transverse components 
    of the charged primary tracks.
    The following pre-cuts were applied prior to the likelihood construction:
    $P_\perp^{mis}<0.8~\GeV/c$, $0.2 \leqslant\theta_h/\pi\leqslant 0.5$ and
    $\alpha/\pi\geqslant 0.8$.

    In Eq.~(\ref{eq:like}) the $P(\vec{\ell}\;|QEL)$ and $P(\vec{\ell}\;|BG)$ are 
    the probabilities for
    signal and background events to have the values of the variables
    $\vec{\ell} = (P_\perp^{mis},\theta_h,\alpha)$. We have found that the DIS and 
    RES probability functions are very similar; therefore we build the likelihood 
    function taking only resonance events for the denominator of Eq.~(\ref{eq:like}).

\begin{figure*}
\begin{center}
\begin{tabular}{cc}
\mbox{\epsfig{file=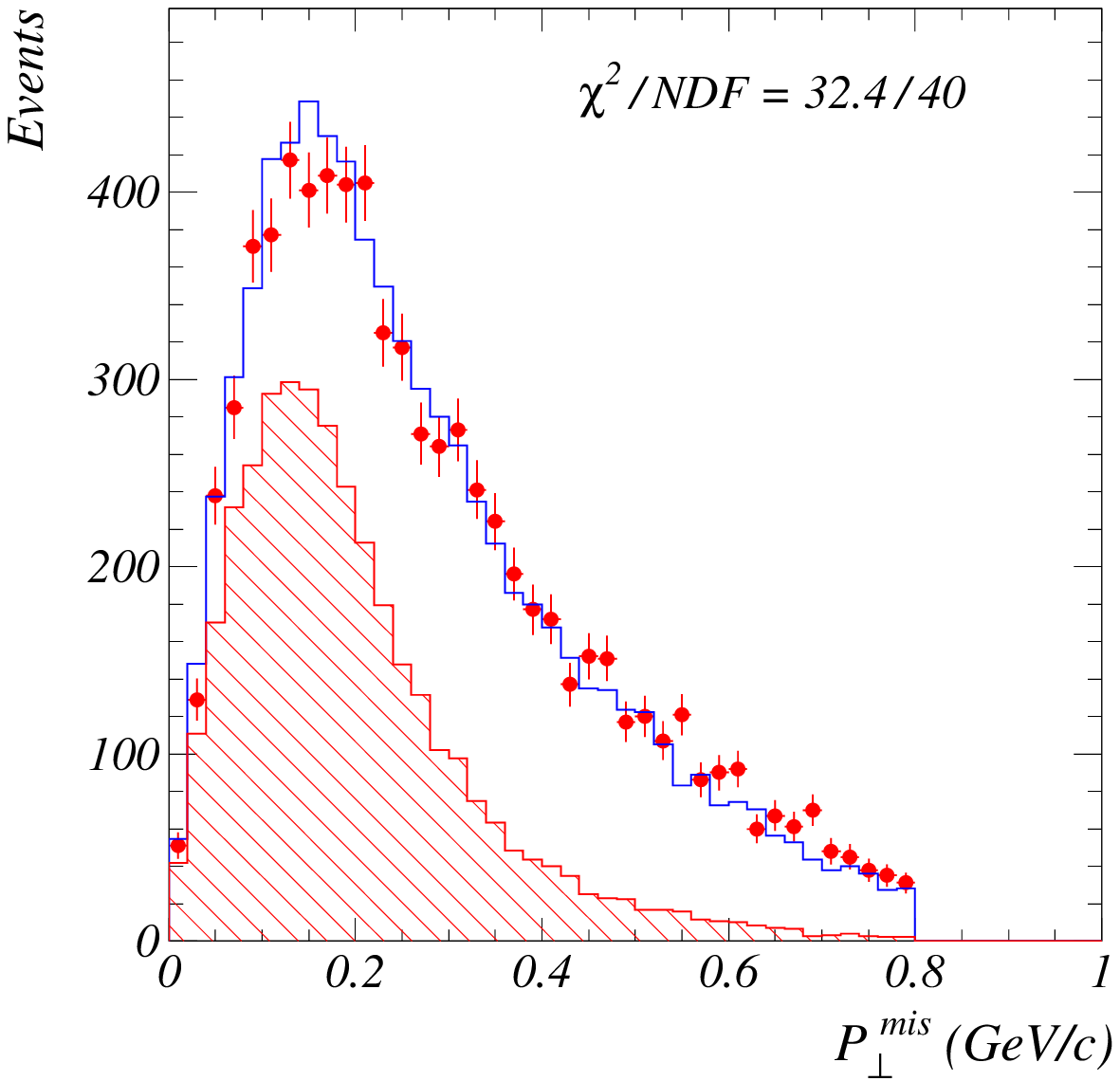,width=0.43\linewidth}} &
\mbox{\epsfig{file=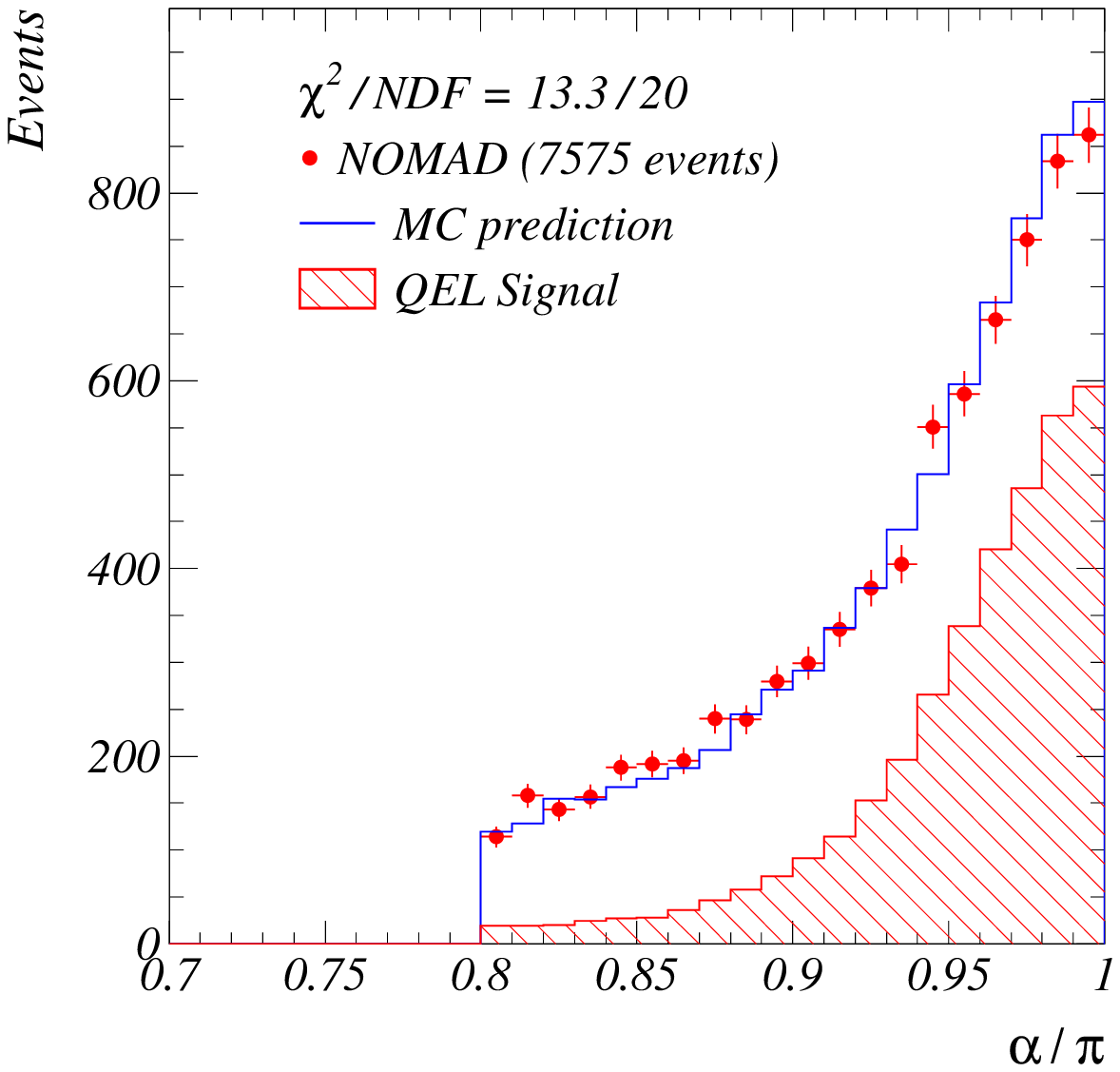,width=0.43\linewidth}} \\
\mbox{\epsfig{file=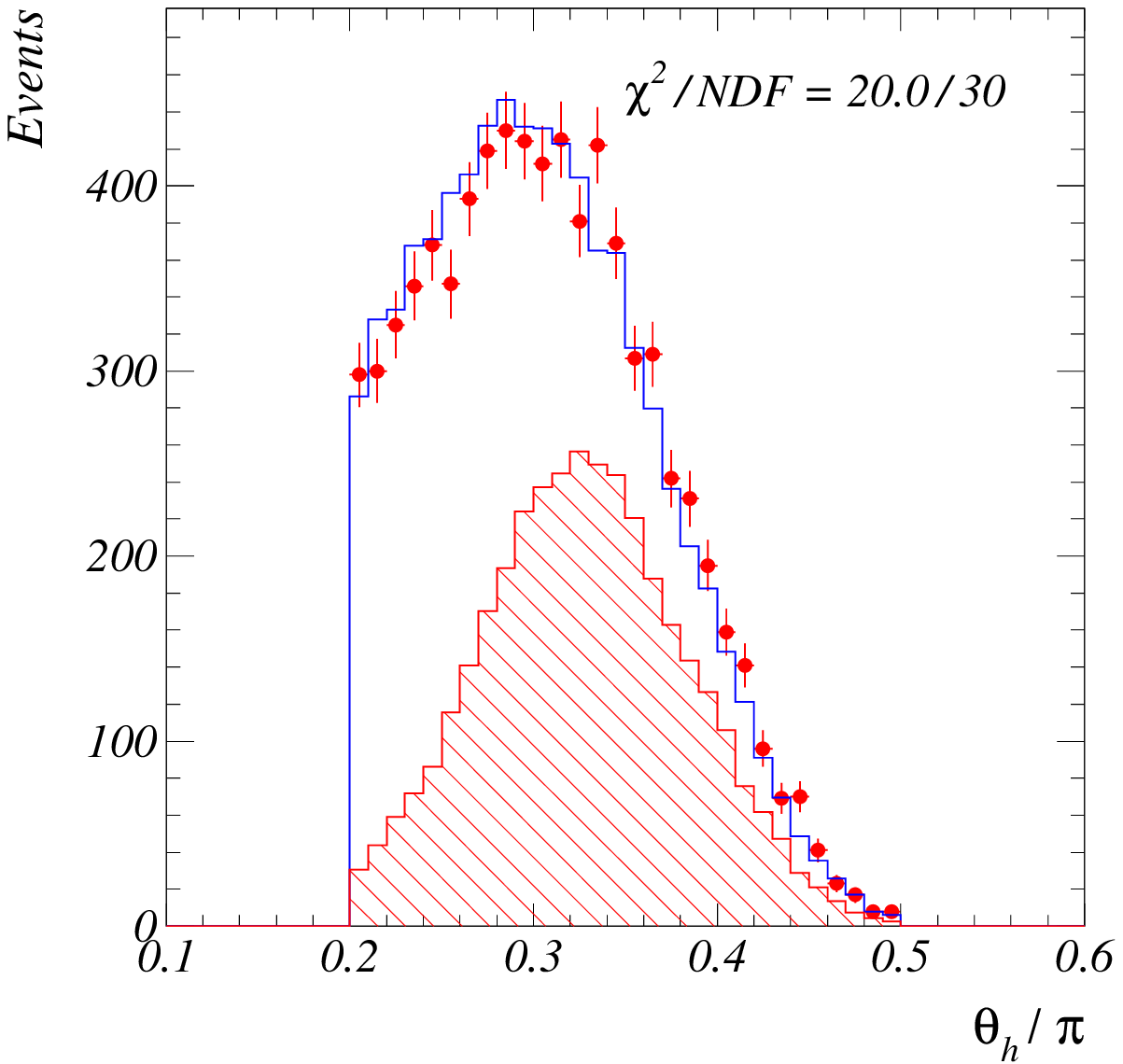,width=0.43\linewidth}} & 
\mbox{\epsfig{file=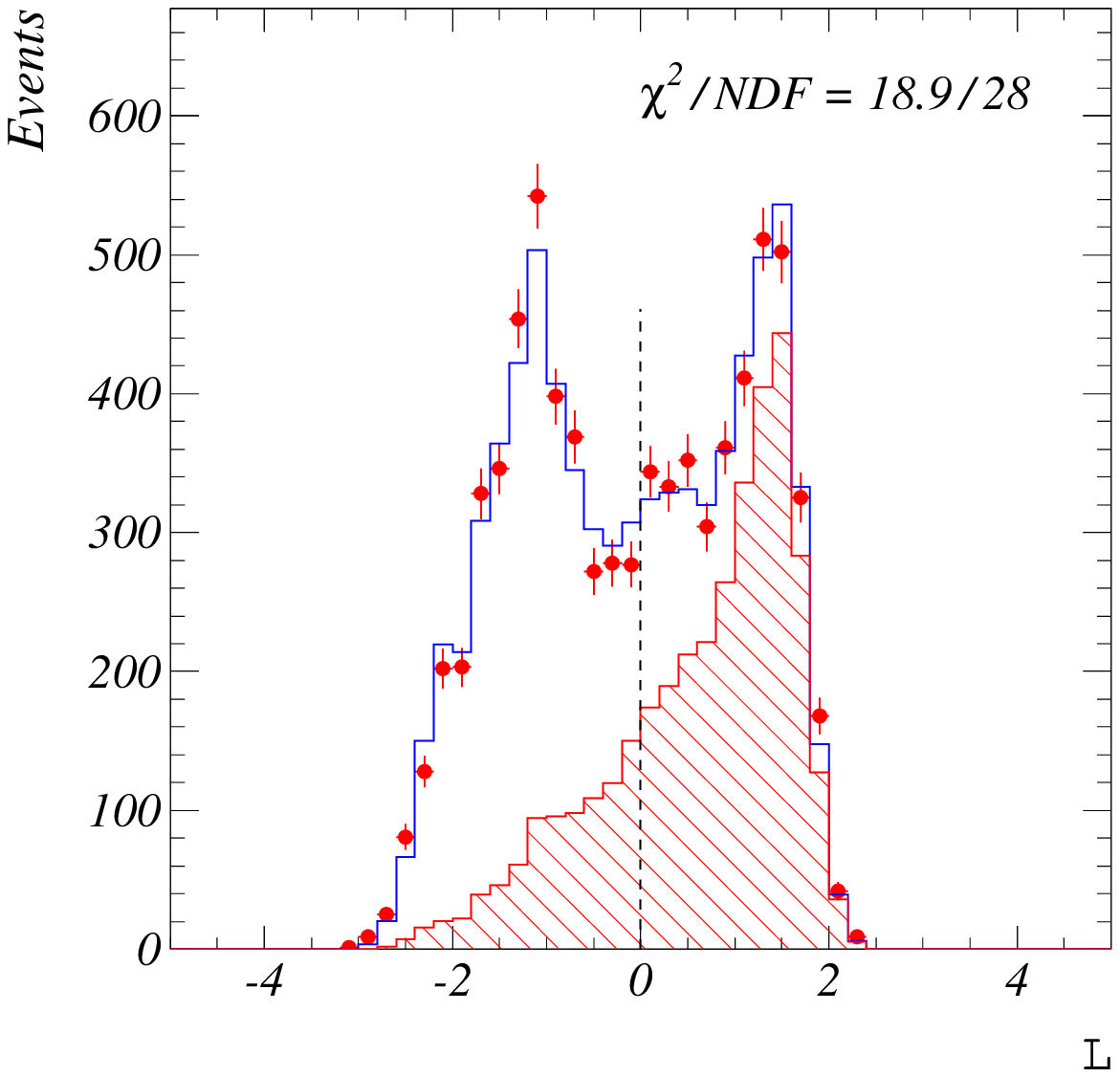,width=0.43\linewidth}} \\
\end{tabular}
\end{center}
\caption{\label{fig:like_var}
The $P_\perp^{mis}$, $\alpha$, $\theta_h$ and likelihood distributions for a mixture 
of QEL, RES and DIS simulated events (histograms) compared to real data
(points with error bars). The MC distributions are normalized to the number 
of events observed in the data. 
}
\end{figure*}

    The comparison of $P_\perp^{mis}$, $\alpha$, $\theta_h$ and $\mathcal{L}$ 
    distributions in the data with the proper mixture of simulated QEL, RES and 
    DIS events is displayed in Fig.~\ref{fig:like_var}. The good agreement 
    observed between MC predictions and experimental data confirms a reasonable 
    understanding of the background contaminations and reconstruction efficiency 
    in our analysis. For example, after the likelihood cut, the respective contributions of QEL, RES, DIS
    and COH given by the cross sections and the efficiencies computed with the
    help of the MC for each process separately are as specified on the
    corresponding $\mathcal{L}>0$ line in Table~\ref{tab:qel_identification}. 
    With the reduction factors for the $\mathcal{L}>0$ cut as given by the MC in the
    various channels, and normalizing the total MC to the data at this last stage, 
    we find a total of 7609 events before the cut, distributed as shown by the 
    figures on the corresponding line. The excess of 34 MC events relative to the
    data, which are necessarily
    mostly background, can be taken as evidence that there is less than 1\%
    excess background in the MC after the pre-cuts. Since the total MC
    background is of the order of 4000 events, the number found is well under
    the expected statistical fluctuations. Therefore, there is no evidence of
    a statistically significant discrepancy.

    In the case of 1-track events, our abilities to suppress background contamination
    are limited since all kinematic variables are expressed in terms of the muon 
    momentum $p_\mu$ and emission angle $\theta_\mu$ with the help of the 
    conservation laws for QEL events. Therefore, the proton reconstructed emission
    angle, Eq.~(\ref{eq:fake_theta_pr}), 
    can be considered as an analog of the likelihood function 
    (see Fig.~\ref{fig:fake_theta_pr}).
    
    The explicit values for the kinematic selection criteria ($\mathcal{L}\geqslant0$ 
    for the 2-track sample and $0.35\leqslant\theta_h/\pi\leqslant 0.5$ for the 1-track 
    sample) were found from the optimization of the sensitivity $SG/\sqrt{SG+BG}$, 
    where $SG$ and $BG$ are the expected numbers of signal and background events 
    in the identified QEL sample.

  \end{itemize}
\end{itemize}

\subsection{$\bar \nu_\mu p \to \mu^+ n$ selection} 

The investigation of antineutrino sample is a much simpler task since these
events are mostly ($\sim 96\%$ of cases) reconstructed as 1-track events (we have
no hits from outgoing neutrons in the drift chambers). Therefore, we require 
identification of the positively charged muon and follow the procedure for 
the 1-track sample discussed above. The only difference is the absence of  
contamination from the inverse muon decay events, so we do not need to apply the
quality cut on the transverse muon momentum.

\vspace{0.3cm}

In Table~\ref{tab:qel_identification} we summarize the information about the selection
of samples with $\nu_\mu n\to \mu^- p$ and $\bar\nu_\mu p\to \mu^+ n$ candidates in
the data. 
The last two columns of this table allow to make checks of compatibility 
between the levels of background in the data and in our simulations 
in a manner similar to what is explained above for the two track sample.

\begin{figure*}
\begin{center}
\mbox{\psfig{file=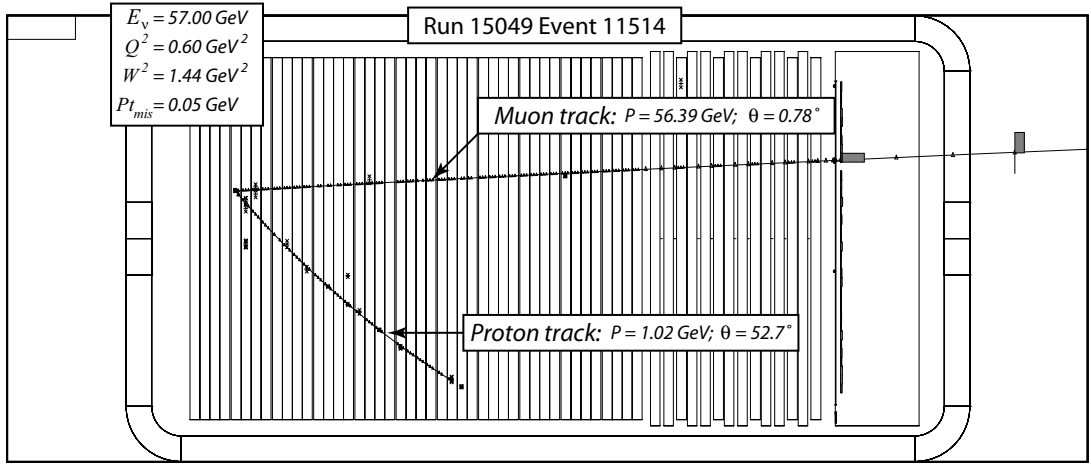,width=0.80\linewidth}}
\end{center}
\caption{\label{fig:qel_events_view} 
A typical example of data event (run 15049 event 11514) 
identified as $\nu_\mu n\to \mu^- p$ in this analysis. 
Long track is identified as muon, short track is assumed to be proton.
}
\end{figure*}

An example of the 2-track event from real data identified as 
$\nu_\mu n\to \mu^- p$ is displayed in Fig.~\ref{fig:qel_events_view}.

\begin{figure*}
\begin{center}
\begin{tabular}{cc}
\mbox{\epsfig{file=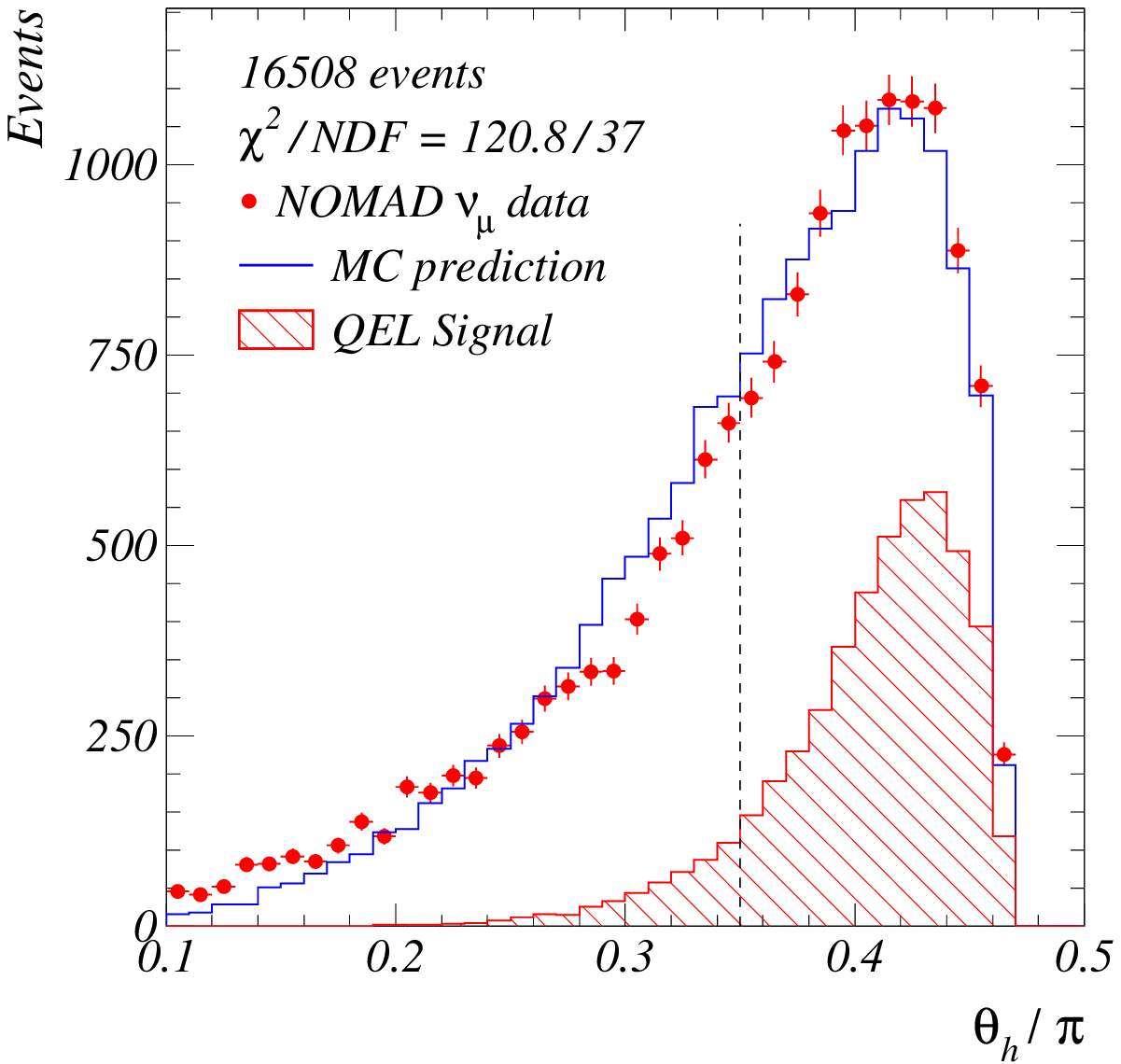,width=0.45\linewidth}} &
\mbox{\epsfig{file=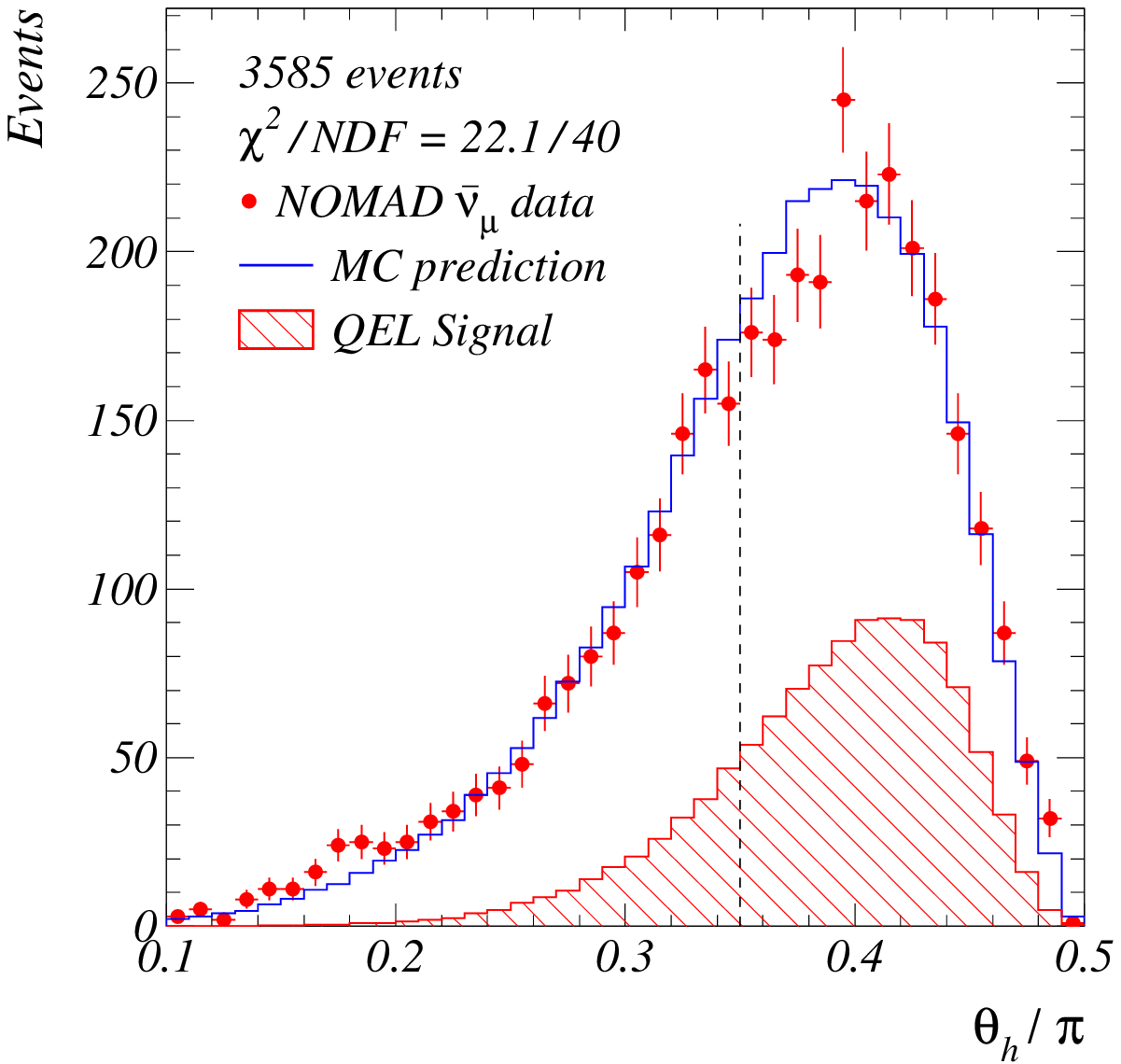,width=0.45\linewidth}} \\
\end{tabular}
\end{center}
\caption{\label{fig:fake_theta_pr}
The $\theta_{h}$ distributions for single track $\nu_\mu$ (left) and $\bar{\nu}_\mu$ (right) samples: comparison of MC distributions (histograms) with the real
data (points with error bars).
}
\end{figure*}

\clearpage

\clearpage
\section{\label{section:QEL_measurement} 
The QEL cross-section and axial mass measurements} 

In this section we describe 
our analysis procedure.\\
The QEL cross-section measurement using normalization
either to the total (DIS) $\nu_\mu$ ($\bar{\nu}_{\mu}$) CC cross-section
or to the inverse muon decay (IMD), $\nu_\mu e^{-}\to \mu^{-} \nu_e$, events 
is first presented in subsection~\ref{section:qel_cs}. 
Afterwards, we describe the procedure used to extract 
the value of the axial mass $M_A$ from the fit of the $Q^2$
distribution. This is the subject of subsection~\ref{section:qel_dcs}.

\subsection{\label{section:qel_cs} The QEL cross-section measurement}
Since there was no precise knowledge of the integrated neutrino flux 
in the NOMAD experiment, we use a different process with a better
known cross-section, recorded at the same time, for the normalization 
of the QEL cross-section.
A similar procedure was often applied in previous neutrino experiments, as 
for example, CERN BEBC~\cite{Allasia:1990uy}. Moreover, the use of another 
process recorded in the same experimental runs allows to reduce 
significantly the systematic uncertainty related to the detector 
material composition. Nevertheless, this auxilliary process must meet two 
requirements: its cross-section should be measured with rather high accuracy 
and the corresponding events can easily be extracted from the full data sample.

Let us divide the investigated interval of neutrino energy into several bins 
and enumerate them with index $i = 1..N_E$. Then, the number of identified QEL 
events in the $i$-th bin with boundaries $[E_i,E_{i+1}]$ is
\begin{equation}
N_i^{dat} = N_i^{bg} + C \sum_{j=1}^{N_E} \varepsilon_{ij}^{qel} \Phi_j\cs{qel}_j  
\label{eq:n_dat}
\end{equation}
where
\begin{equation*}
\Phi_i  = \int_{E_i}^{E_{i+1}} \Phi(E)\, dE, \quad 
\sum_{i=1}^{N_E} \Phi_i = 1
\end{equation*}
and 
\begin{equation*}
\cs{qel}_i = \frac{1}{\Phi_i} \int_{E_i}^{E_{i+1}} \sigma_{qel}(E)\Phi(E)\, dE \\
\end{equation*}

Coefficient $C$ accumulates the absolute neutrino flux and the number of target 
nucleons. The matrix element $\varepsilon_{ij}^{qel}$ is the probability that
the reconstructed neutrino energy $E_\nu$ of a QEL event falls into the $i$-th bin, 
while the simulated energy actually belongs to the $j$-th bin. 

The expected background contamination is
\begin{equation}
N^{bg}_i = C\, (\varepsilon^{res}_i \cs{res} + \varepsilon^{dis}_i \cs{dis})
\label{eq:n_bg}
\end{equation}
where we use the definition of Eq.~(\ref{eq:acs}) for $\cs{bg}$; $\varepsilon^{bg}_i$ 
denotes the renormalized energy distribution in BG events passing the QEL 
identification procedure: 
\begin{equation}
\sum_{i=1}^{N_E}\varepsilon^{bg}_i = \varepsilon^{bg} = N^{bg}_{rec}/N^{bg}_{sim}
\label{eq:eff_bg}
\end{equation}
here $N^{bg}_{sim}$ and $N^{bg}_{rec}$ are the number of MC events simulated 
and identified as QEL in the chosen detector FV.

Similar equations can be written for any other process recorded in the
same detector FV. If we identify $N_0$ events of a process, whose
flux averaged cross section in an energy interval containing these
events is $\sigma_0$, we can write
\begin{equation*}
N_0 = C\, \Phi_0 \sigma_0 
\end{equation*}
where $\Phi_0$ is the relative part of the neutrino flux belonging to 
the same energy interval.  
(we assume that $N_0$ is background subtracted and efficiency corrected).

We can now get rid of $C$ and write the final equation for $\cs{qel}_i$:
\begin{align}
\cs{qel}_i & = \frac{1}{\Phi_i}\sum_{j=1}^{N_E} (\varepsilon^{-1}_{qel})_{ij} 
\times \nonumber \\
& \qquad \left[N^{dat}_j\frac{\Phi_0\sigma_0}{N_0} - \varepsilon^{res}_j \cs{res} 
  -\varepsilon^{dis}_j\cs{dis} \right]
\label{eq:cs_qel_i}
\end{align}

Numerical values for $\cs{res}$ and $\cs{dis}$ are given in  
Table~\ref{table:cs_all}. The efficiencies $\varepsilon^{qel}_{ij}$, 
$\varepsilon^{res}_i$ and $\varepsilon^{dis}_i$ should be estimated with the 
help of the MC simulation for QEL, RES and DIS samples separately;
the factor $\Phi_0\sigma_0/N_0$ comes from the auxilliary process used for 
normalization.

Let us note that the smearing of the reconstructed neutrino energy is taken 
into account in Eq.~(\ref{eq:cs_qel_i}) by the inverse matrix 
of QEL efficiencies. 

Equation~(\ref{eq:cs_qel_i}) can also be applied to the entire energy interval.
In this case, we can use the usual notations for efficiencies as 
in Eq.~(\ref{eq:eff_bg}). From the measured $\cs{qel}$ we calculate the axial 
mass $M_A$ by using the Smith and Moniz formalism (see Fig.~\ref{fig:acs_qel}).

In the following subsections, we investigate the DIS and IMD processes which can 
both be used for the QEL cross-section normalization as just described.

Possible sources of systematic errors in our analysis procedure are discussed
in Section~\ref{section:systematics}.  

\subsubsection{\label{section:dis_norm} Selection of DIS events}

\begin{table*}[t]
\caption{\label{table:dis_identification}
Selection of the DIS events in $\nu_\mu$ and $\bar{\nu}_\mu~CC$ samples. 
Total efficiency (in \%), expected purity of selected events (in \%), 
theoretical prediction for $\cs{dis}$, observed $N_{dat}$ and corrected $N_0$ 
number of events in experimental data are given for each variant of DIS selection 
described above. 
\hfill}
\begin{tabular}{r >{\hspace{10pt}}rrr >{\hspace{10pt}}rrr}
\hline\noalign{\smallskip}
& \multicolumn{3}{c}{$\nu_\mu$ sample} & \multicolumn{3}{c}{$\bar\nu_\mu$ sample} \\
variant of DIS selection         &        1 &        2 &        3 &        1 &       2 &       3 \\
\noalign{\smallskip}\hline\noalign{\smallskip}
efficiency                       &    82.95 &    86.84 &    88.52 &    75.46 &   81.40 &   83.20 \\
purity                           &    97.10 &    98.62 &    99.62 &    71.48 &   72.57 &   73.95 \\
$N_{dat}$, events                & 676702.0 & 267517.0 & 276018.0 &  17744.0 &  7996.0 &  8500.0 \\
\noalign{\smallskip}\hline\noalign{\smallskip}
$N_0$, events                    & 792162.0 & 303790.7 & 310617.3 &  16807.1 &  7128.6 &  7553.4 \\
relative flux $\Phi_0$           &        1 &    0.144 &    0.144 &        1 &   0.106 &   0.106 \\
$\cs{0}$, $10^{-38}~\cm^2$       &   16.643 &   44.876 &   46.069 &    4.876 &  20.124 &  21.999 \\
$C^{-1}$, $10^{-43}~\cm^2$       &    2.101 &    2.127 &    2.136 &   29.012 &  29.924 &  30.872 \\
\noalign{\smallskip}\hline
\end{tabular}
\end{table*}

The phenomenology of neutrino DIS is well developed. Experimental data are in
rather good agreement with theoretical predictions. The charged current neutrino
DIS is an inclusive process and for its selection from the data sample, 
the following criteria are enough:
\begin{itemize}
\item {\em Fiducial volume cut.} The primary vertex should be in the same FV 
  as that defined for the QEL events, see Eq.~(\ref{eq:fv});
\item {\em Muon identification and Topology cut.} 
  At least two charged tracks should originate from the primary vertex; one of 
  them should be identified as a muon ($\mu^-$ in the case of $\nu_\mu$~CC and
  $\mu^+$ for $\bar{\nu}_\mu$~CC);
\item {\em Background suppression.}
  The third criterion is used to avoid contributions from the QEL and RES events.
  We have checked three different possibilities for it:
  \begin{enumerate}
  \item The total visible energy in the event should be $E_\nu \leqslant 300~\GeV$ 
    and the reconstructed hadronic mass $W \geqslant 1.4~\GeV$; in this case the 
    computation of $\cs{dis}$ has been done for GRV98-LO PDF model according 
    to the prescriptions in~\cite{Kuzmin:2005bm}.
  \item We keep the requirement for the reconstructed hadronic mass 
    ($W \geqslant 1.4~\GeV$) but reduce the neutrino energy region to 
    $40 \leqslant E_\nu \leqslant 200~\GeV$; theoretical calculation of 
    $\cs{dis}$ is also done with the help of~\cite{Kuzmin:2005bm}.
  \item Using the same neutrino energy interval as in 2.  
    ($40 \leqslant E_\nu \leqslant 200~\GeV$), we remove the cut on the reconstructed 
    hadronic mass $W$. In this case, we take the total CC neutrino-nucleon 
    cross-section to be:  
    \footnote{The CHORUS measurement for the $CH_2$ target~\cite{Chorus} 
      is consistent with this value.} 
    \begin{align*}
      \sigma^{tot}_\nu(E_\nu)/E_\nu & =
         (0.677\pm 0.014)\times 10^{-38}~\cm^2/\GeV \\
      \sigma^{tot}_{\bar{\nu}}(E_\nu)/E_\nu & = 
         (0.334\pm 0.008)\times 10^{-38}~\cm^2/\GeV
    \end{align*}
    (PDG average~\cite{Eidelman:2004wy}). The calculated $\cs{tot}$ should be 
    corrected due to the fact that NOMAD target is slightly non-isoscalar.
  \end{enumerate}
\end{itemize}

The numerical results of the DIS events selection can be found in 
Table~\ref{table:dis_identification}. For the QEL cross-section normalization
we use results obtained with the last method (PDG based) as having the most solid ground.
Thus, the final normalization is performed 
to the total $\nu_\mu$ ($\bar{\nu}_{\mu}$) CC cross-section.
We also checked that this normalization is consistent with two previous calculations 
based on approach from~\cite{Kuzmin:2005bm} within $1.6\%$ ($5.9\%$) 
for $\nu_\mu$ ($\bar{\nu}_\mu$) CC sample.

\subsubsection{\label{section:imd} Selection of inverse muon decay events}

Inverse muon decay $\nu_\mu e^{-}\to \mu^{-} \nu_e$ is a purely
leptonic process, which is well known both on theoretical and experimental grounds. 
Its cross-section in the Born approximation is:
\begin{equation}
\sigma_{imd}(E_\nu) = \sigma_{as} 
E_\nu \left(1-\frac{m_\mu^2}{2m_e E_\nu}\right)^2
\label{cs_imd}
\end{equation}
The numerical value of the constant $\sigma_{as}$ calculated in the framework of 
the Standard Model was found to be in good agreement with experimental 
measurements~\cite{CHARM_II}:
\begin{equation}
\sigma_{as} = \frac{2m_e G_F^2}{\pi} = 1.723\times 10^{-41}~\cm^2~\GeV^{-1}
\end{equation}

The number of IMD events $N_0$ is proportional to its flux averaged 
cross-section from Eq.~(\ref{eq:acs}):
\begin{equation}
\cs{imd} = 1.017\times 10^{-40}~\cm^2
\end{equation}
and expected to be at least 650 times smaller than the number of DIS events.

To select the IMD events we require:
\begin{itemize}
\item the primary vertex should be in the same fiducial volume as that used for 
identified QEL events, see Eq.~(\ref{eq:fv});
\item there is only one negatively charged track originating from the primary vertex;
  it should be identified as a muon;
\item there are no veto chamber hits in the vicinity of the intersection point
  of the extrapolated muon track and the first drift chamber (quality cut, the same as
  for 1-track events from the QEL sample);
\item the muon energy is above the threshold:
  \begin{equation}
    E_\mu \geqslant \frac{m_\mu^2+m_e^2}{2m_e} = 10.93~\GeV
  \end{equation}
\item the transverse momentum $p_{\perp}$ of the muon produced in IMD event is very 
    limited by kinematics: $p_{\perp}^2 \leqslant 2m_e E_\mu$.
\end{itemize}

\begin{figure}
\begin{center}
\begin{tabular}{c}
\mbox{\epsfig{file=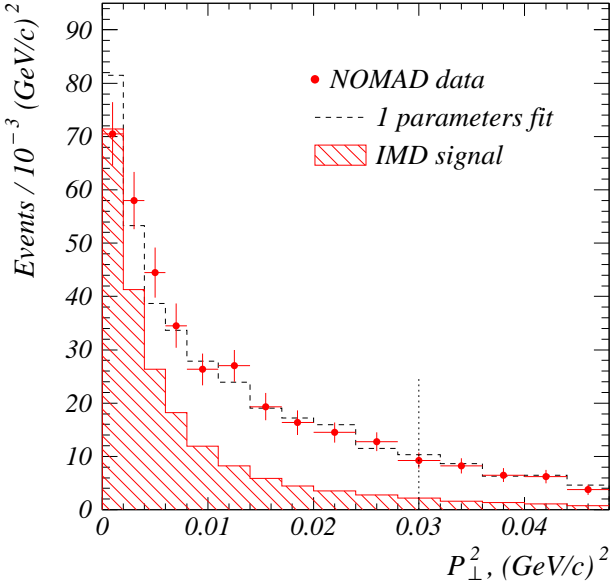,width=0.95\linewidth}}
\end{tabular}
\end{center}
\caption{\label{fig:imd_rd}
Inverse Muon Decay: NOMAD experimental data, the non-equidistant $p_{\perp}^2$ 
distribution.
}
\end{figure}

In this sample the contamination from the reaction
$\bar \nu_e e \rightarrow \mu^- \bar \nu_\mu$ is estimated to be at the
level of $\sim 10^{-3}$, e.g. well below 1 event, since the ratio of
the fluxes $\bar \nu_e / \nu_\mu$ is 0.0027~\cite{Astier:2003rj}
while the ratio of the cross-sections is
$\sigma (\bar \nu_e e \rightarrow \mu^- \bar \nu_\mu) /
\sigma (\nu_\mu e \rightarrow \mu^- \nu_e) \approx 1/3$.

We determine the number of signal events $N_{imd}$ from the fit of 
the $p^2_{\perp}$ distribution to experimental data with the function $F(p^2_{\perp})$:
\begin{equation}
F(p^2_{\perp}) = N_{imd}F_{imd}(p^2_{\perp})
 +\left[N_{dat}-N_{imd}\right]F_{bg}(p^2_{\perp})
\label{eq:fit_imd}
\end{equation}
where $F_{imd}$ and $F_{bg}$ are the normalized MC expectations for signal and background
$p^2_{\perp}$ distributions; $N_{dat}$ denotes the 
number of events in real data which passed all selection criteria. 

The QEL events are now playing the role of the most important background for the 
IMD selection. However, the contaminations from the RES and DIS events cannot be
neglected since they distort the shape of the $p_\perp^2$ distribution. As usual,
the relative contribution of each process to the expected background is proportional 
to the corresponding efficiency and flux averaged cross-section 
(see Table~\ref{table:cs_all}).

The expression (\ref{eq:fit_imd}) contains only one free parameter $N_{imd}$,
which is the number of observed IMD events. Finally,
for $p^2_{\perp}<0.03~(\GeV/c)^2$ interval 
we find $N_{imd} = 436.0\pm 28.5$  with the quality of the fit 
$\chi^2/NDF = 0.89$ (see Fig.~\ref{fig:imd_rd}). Taking into account that the 
selection efficiency for the IMD events is $87.8\%$ we report the total number 
of IMD events $N_0$, which can be used for the QEL normalization:
\begin{equation}
N_0 = 496.6 \pm 32.5
\end{equation}

The relative error for $\sigma_0/N_0$ in the IMD case is about $7\%$ (due to the
small statistics of the IMD sample). Nevertheless the normalization factor itself,
$C^{-1} = 2.048 \times 10^{-43}~\mbox{cm}^2$, is in agreement 
(within $\sim 4\%$) with the evaluation based on the DIS sample 
(see Table~\ref{table:dis_identification}).

The use of the IMD process for the normalization is an interesting
independant cross-check of the absence of problems in our procedure.
In particular, it allows to verify that there are no effects arising
from possible trigger inefficiencies in the selection of neutrino
events consisting of a single muon going through the trigger planes.

\subsection{\label{section:qel_dcs} Axial mass measurement from the $Q^2$ distribution}

To extract the axial mass from the $Q^2$ distribution the experimental data  
are fitted to the theoretical predictions using a standard $\chi^2$ method. 
We bin the events in two variables $Q^2$ and $E_\nu$ (in the case of a single
$E_\nu$ interval our procedure can be considered as the usual 1-dimensional fit)
\footnote{In practice it is convenient to use dimensionless variables $(a,b)$ 
instead of $(E_\nu,Q^2)$. Then, $E_\nu = E_\nu^{min} + a(E_\nu^{max}-E_\nu^{min})$ 
and  $Q^2 = Q^2_{min}(E_\nu) + b[Q^2_{max}(E_\nu)-Q^2_{min}(E_\nu)]$. 
So, $a,b\in[0,1]$.}.

Let us enumerate bins with index $i = 1..N_B$; bin $i=N_B+1$ contains events which fall outside of the investigated $(E_\nu,Q^2)$ region. It is convenient to define
boundaries in such a way that each bin with $i = 1..N_B$ contains approximately 
the same number of experimental events passing all identification criteria.

A minimization functional is:
\begin{equation}
\chi^2(M_A) = \sum_{i=1}^{N_B} 
              \frac{\left[N_i^{dat}-N_i^{th}{(M_A)}\right]^2}{N_i^{dat}}
\label{eq:chi2}
\end{equation}
where $N_i^{dat}$ is the number of events in the $i$-th bin of the non-weighted 
experimental distribution, while $N_i^{th}$ is a superposition of the normalized 
MC background $N^{bg}_i$ and the expected QEL signal:  
\begin{equation}
N_i^{th}(M_A) = N_i^{bg} + 
  C \sum_{j=1}^{N_B+1} \varepsilon^{qel}_{ij} \Phi_j \dcs{qel}_j
\end{equation}

This equation is similar to Eq.~(\ref{eq:n_dat}), $N_i^{bg}$ being defined in
the same way as in Eq.~(\ref{eq:n_bg}); $\varepsilon_{ij}^{qel}$ is the probability that
a QEL event simulated in the $j$-th bin is reconstructed in the $i$-th bin. 
The QEL scattering dynamics is described by the following term:
\begin{align}
& \dcs{qel}_i = \frac{1}{\Phi_i}\int_{\Omega_i} 
  \frac{d\sigma}{dQ^2}(E,Q^2,M_A) \Phi(E)\,dE dQ^2 \\
& \left.\Phi_i\dcs{qel}_i\,\right|_{i=N_B+1} = 
  \cs{qel}-\sum_{j=1}^{N_B}\Phi_j\dcs{qel}_j
\end{align}
here $\Omega_i$ denotes the $(E_\nu,Q^2)$ interval, which corresponds to the 
$i$-th bin; $d\sigma/dQ^2$ is the differential QEL cross-section on bound target
nucleon 
(see formulae in~\cite{Kuzmin:2007kr})

The coefficient $C$ can be defined in either of two ways:
\begin{enumerate}
\item the $N_i^{th}$ distribution is normalized to the total number of events
in the experimental data:
\begin{equation}
\sum_{i=1}^{N_B}N_i^{th} = \sum_{i=1}^{N_B}N_i^{dat}
\end{equation}
In this case, the proposed method should be sensitive only to the shape of the 
distribution but not to the absolute number of identified events (contrary to 
the $M_A$ measurement from the total QEL cross-section).
\item $C$ is defined in the same way as for the total QEL cross-section measurement,
i.e.~we use another process (DIS) for normalization:
\begin{equation}
C = \frac{N_0}{\Phi_0\sigma_0}
\end{equation}
If we sum over the $Q^2$ variable for the investigated $(E_\nu,Q^2)$ 
interval, finding the $M_A$ parameter from Eq.~(\ref{eq:chi2}) becomes nothing else 
than the numerical resolution of Eq.~(\ref{eq:n_dat}). Therefore, this variant of the fit can 
be considered as a simultaneous fit of the total and differential cross-sections; 
henceforth, we shall refer to it as $\tlD$ fit.
\end{enumerate}

\begin{figure}
\begin{center}
\mbox{\epsfig{file=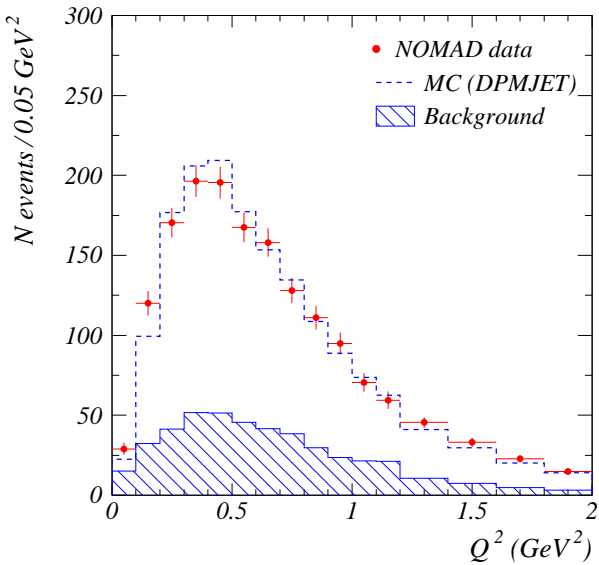,width=0.9\linewidth}}
\end{center}
\caption{\label{fig:q2_rec}
The $Q^2$ distributions in identified QEL events.
}
\end{figure}

Fig.~\ref{fig:q2_rec} presents a comparison of the reconstructed $Q^2$ distribution
with our MC prediction.
The expected background contamination is also shown.

We can now apply the proposed methods to experimental data and 
measure the QEL cross-section and axial mass $M_A$. The numerical results 
are reported in Section~\ref{section:results}, while the discussion of the 
corresponding uncertainties is presented in the next section.

\begin{table*}
\caption{\label{table:systematics}
The relative systematic uncertainties (in \%) of the QEL cross section $\cs{qel}$ and 
axial mass $M_A$, measured in $\nu_\mu n\to \mu^- p$ and 
$\bar{\nu}_\mu p\to \mu^+ n$ reactions.
\hfill
}
\begin{tabular}
{r l >{\hspace*{10pt}}rrr >{\hspace*{10pt}}rr}
\hline\noalign{\smallskip}
& {\em Source} & $\cs{qel}_{\nu_\mu}$ 
& $M_A$ {\em from} $\cs{qel}_{\nu_\mu}$ 
& $M_A$ {\em from} $d\sigma_\nu/dQ^2$ 
& $\cs{qel}_{\bar{\nu}_\mu}$ & $M_A$ {\em from} $\cs{qel}_{\bar{\nu}_\mu}$ \\
\noalign{\smallskip}\hline\noalign{\smallskip}
1 & QEL identification procedure: \\
  & likelihood or $\theta_h$ cut  &    3.5 &       3.2 &        2.4 &    4.3 &     4.2 \\
  & $\varphi_\mu$ cut    &    0.8 &       0.7 &        0.3 &     -- &      -- \\
  & $P_\perp^{mis}$, $\alpha$ and $\theta_h$ precuts &
                                       0.4 &       0.4 &        0.4 &     -- &      -- \\
2 & $\delta(\sigma_{dis})$        &    2.9 &       2.6 &        0.2 &    4.2 &     4.2 \\
3 & $\delta(\sigma_{res})$        &    4.0 &       3.6 &        0.6 &    7.6 &     7.4 \\
4 & nuclear reinteractions        &    1.8 &       1.6 &        6.5 &     -- &      -- \\
5 & shape of neutrino spectrum    &    0.2 &       0.2 &        0.1 &    0.9 &     0.9 \\
6 & NC contribution               & $<0.1$ &    $<0.1$ &         -- &    1.1 &     1.1 \\
7 & muon misidentification        & $<0.1$ &    $<0.1$ &         -- &    1.0 &     1.0 \\
8 & coherent pion production      & $<0.1$ &    $<0.1$ &     $<0.1$ &    1.1 &     1.1 \\
\noalign{\smallskip}\hline\noalign{\smallskip}
  & total                         &    6.5 &       5.9 &        7.0 &    9.9 &     9.5 \\
\noalign{\smallskip}\hline
\end{tabular}
\end{table*}

\section{\label{section:systematics} Systematic uncertainties}

We have studied several sources of systematic uncertainties, which are important
for the measurement of the total QEL cross-section and axial mass parameter. 
They are listed below:
\begin{enumerate}
\item identification of QEL events; we vary the selection criteria 
  within reasonable limits ($\mathcal{L}>0\pm 0.4$ for 2-track sample
  and $\theta_h/\pi>0.35\pm 0.03$ for 1-track sample).

  The final result is found to be practically insensitive to the 
  exact positions of the muon azimuth $\varphi_\mu$ cut and additional 
  requirements for the $P_\perp^{mis}$, $\alpha$ and $\theta_h$ variables:
  e.g. in the $\nu_\mu$ analysis a more strict cut $0.1\pi < \varphi_\mu < 0.9\pi$ leads to 
  $0.8\%$ variation in the measured cross section while a change in the 
  pre-cuts to $P_\perp^{mis}<0.9~\GeV$, $\alpha/\pi> 0.75$ and $0.18\pi<\theta_h$
  leads to an uncertainty of $0.4\%$.

\item uncertainty in the total (mainly DIS) charged current muon neutrino
  cross-section, which enters both in the normalization
  factor $\sigma_0/N_0$ and in the subtraction of the corresponding DIS background
  (the experimental error on $\cs{dis}$ is $2.1\%$ for $\nu_\mu$ CC and 
  $2.4\%$ for $\bar{\nu}_\mu$ CC);
\item uncertainty in the RES cross section, which determines the contamination
  admixture of the single resonant pion events in the identified QEL sample (we assume 
  $10\%$ error on $\cs{res}$ both for neutrino and antineutrino cases, 
  see e.g.~\cite{Furuno:2003ng});
\item FSI interactions (we vary $\tau_0$ and $\alpha_{mod}^F$ DPMJET parameters
  for fixed $M_A^{mc} = 1.03~\GeV$);
\item uncertainty in the neutrino flux shape (the relative errors for each 
  $E_\nu$ bin were taken from~\cite{Astier:2003rj});
\item neutral current admixture (we assume $5\%$ error for the corresponding 
  cross section, which can be found in Table~\ref{table:cs_all});
\item charge misidentification of the primary lepton (reconstructed $\nu_\mu$ CC
  event is classified as $\bar{\nu}_\mu$ CC and vice-versa);
\item contamination from coherent pion production (see 
  subsection~\ref{section:coherent_pion_production}).
\end{enumerate}

In Table~\ref{table:systematics} we present our numerical estimations for systematic
uncertainties (in the case of $\nu_\mu$ scattering, systematic errors were calculated for
the mixture of 1-track and 2-track subsamples). One can see 
that the most important contributions come from the QEL identification procedure and 
from the uncertainty on the non-QEL processes contribution to the selected sample 
of signal events.

The nuclear reinteractions (FSI effect) significantly affect the neutrino sample only
(see Table~\ref{table:results_cc}), while in the antineutrino case the influence of 
the nuclear reinteractions is expected to be negligible. For $\nu_\mu$ scattering, 
the cross-sections can be calculated separately for both the 1-track and 2-track 
subsamples of identified QEL events or for their mixture. We can then compare the  
results and choose whichever one has the minimal total error. In our case it was 
obtained for the combined 1-track and 2-track sample, which was found to be 
almost insensitive to the variation of DPMJET parameters (see Section~\ref{section:results} 
for explanations). 

The uncertainty on the shape of the (anti)neutrino spectrum is
important for the measurement of $\sigma_{qel}$ as a function of neutrino
energy $E_\nu$. But it does not affect both the flux averaged cross section  $\cs{qel}$ 
and the $M_A$ extraction from the $Q^2$ distribution.

The uncertainty due to the primary lepton misidentification
and neutral currents comes into play through the subtraction of the corresponding 
background from the selected DIS sample, that is, from the normalization factor. The 
admixture of those events into the identified QEL events is negligible.

\section{Results}
\label{section:results}

\begin{figure*}
\begin{center}
\begin{tabular}{c}
\mbox{\psfig{file=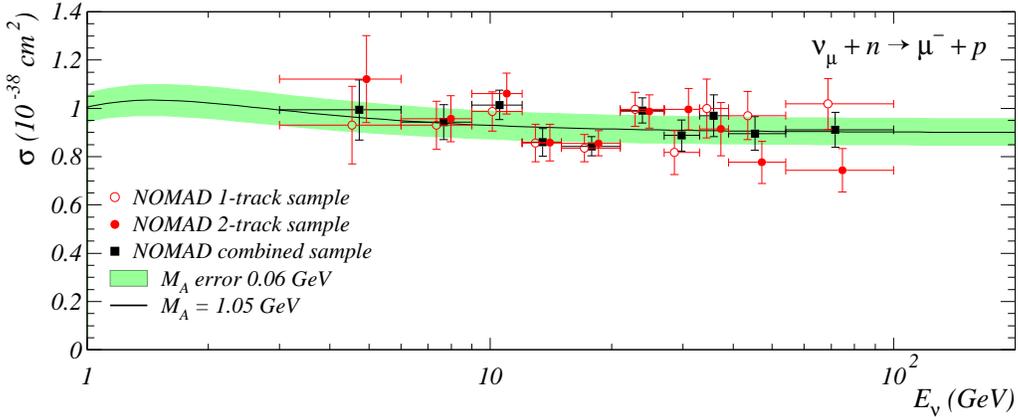,width=0.75\linewidth}} 
\end{tabular}
\end{center}
\caption{\label{fig:nomad_cc_carbon_t}
Comparison of our $\cs{qel}_{\nu_\mu}$ measurements as a function of
the neutrino energy in the 1-track and 2-track 
subsamples (for the best parameter $\tau_0 = 1.0$) with the final 
$\cs{qel}_{\nu_\mu}$ values measured using the full event sample, 
see Table~\ref{table:nomad_qel_cc}.
}
\end{figure*}

\begin{figure*}
\begin{center}
\begin{tabular}{c}
\mbox{\psfig{file=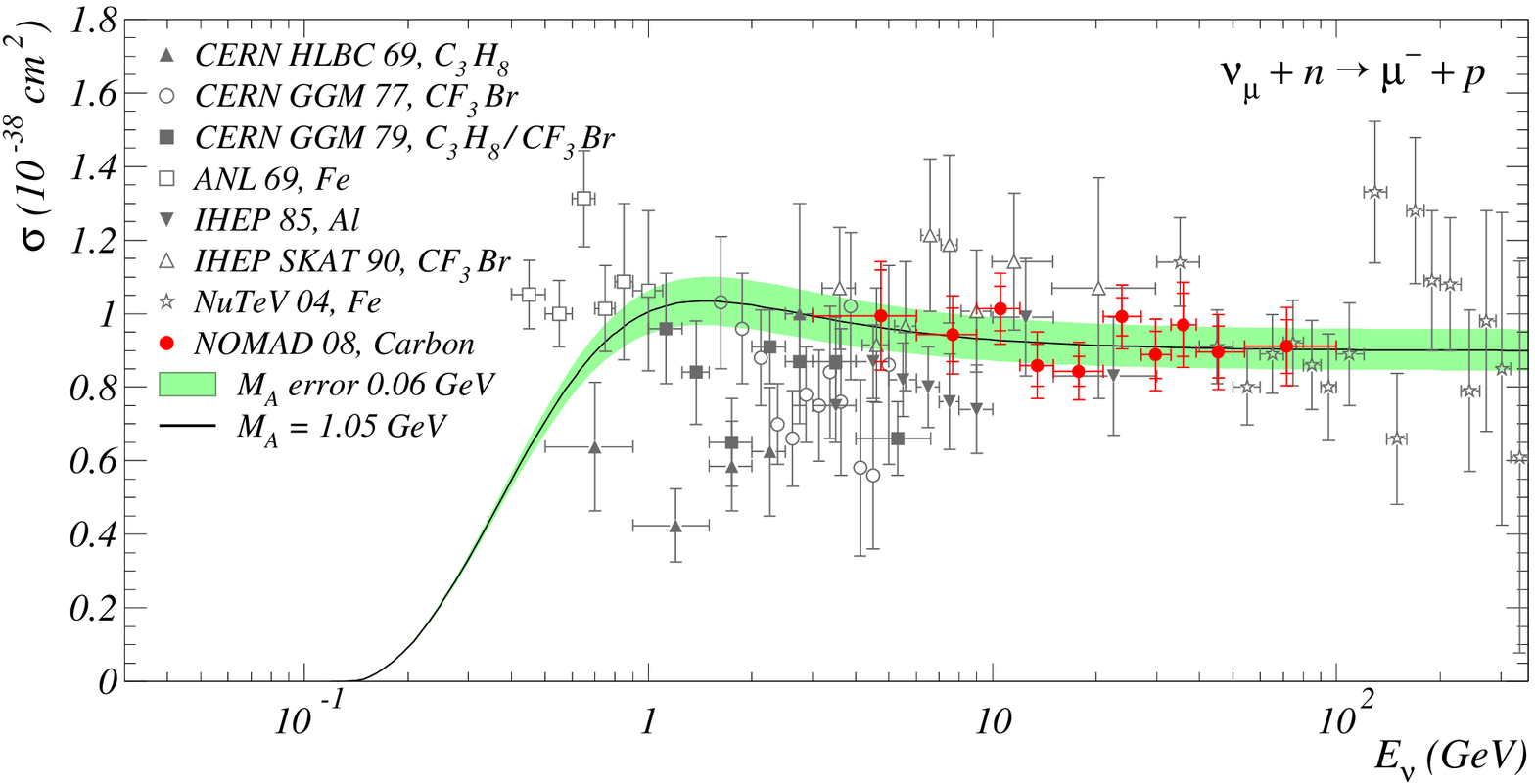,width=0.75\linewidth}} 
\end{tabular}
\end{center}
\caption{\label{fig:nomad_cc_nucl}
Comparison of NOMAD $\cs{qel}_{\nu_\mu}$ 
measurements with previous experimental data on 
$\nu_{\mu}$~scattering off heavy nuclei 
(ANL~69 (Spark-chamber)~\cite{Kustom:1969dh}, 
NuTeV~04 (FermiLab)~\cite{Suwonjandee:04}, 
CERN~HLBC~69 (CERN, Heavy Liquid Bubble Chamber)~\cite{Budagov:1969bg}, 
CERN~GGM~77 (CERN, Gargamelle BC)~\cite{Bonetti:1977cs}, 
CERN~GGM~79~\cite{Pohl:1979zm}, 
IHEP~85 (IHEP, spark-chamber)~\cite{Belikov:1983kg}, 
IHEP~SCAT~90 (IHEP, BC)~\cite{Brunner:1989kw}).
The solid line corresponds 
to the $M_A$ value obtained in the NOMAD experiment, the error band 
takes into account both statistical and systematic uncertainties of 
the present analysis. Nuclear effects are included into calculations 
according to the relativistic Fermi gas model by Smith and
Moniz~\cite{Smith:1972xh} for Carbon with binding energy $E_b = 25.6~\MeV$ 
and Fermi momentum $P_F = 221~\MeV/c$. 
}
\begin{center}
\begin{tabular}{c}
\mbox{\psfig{file=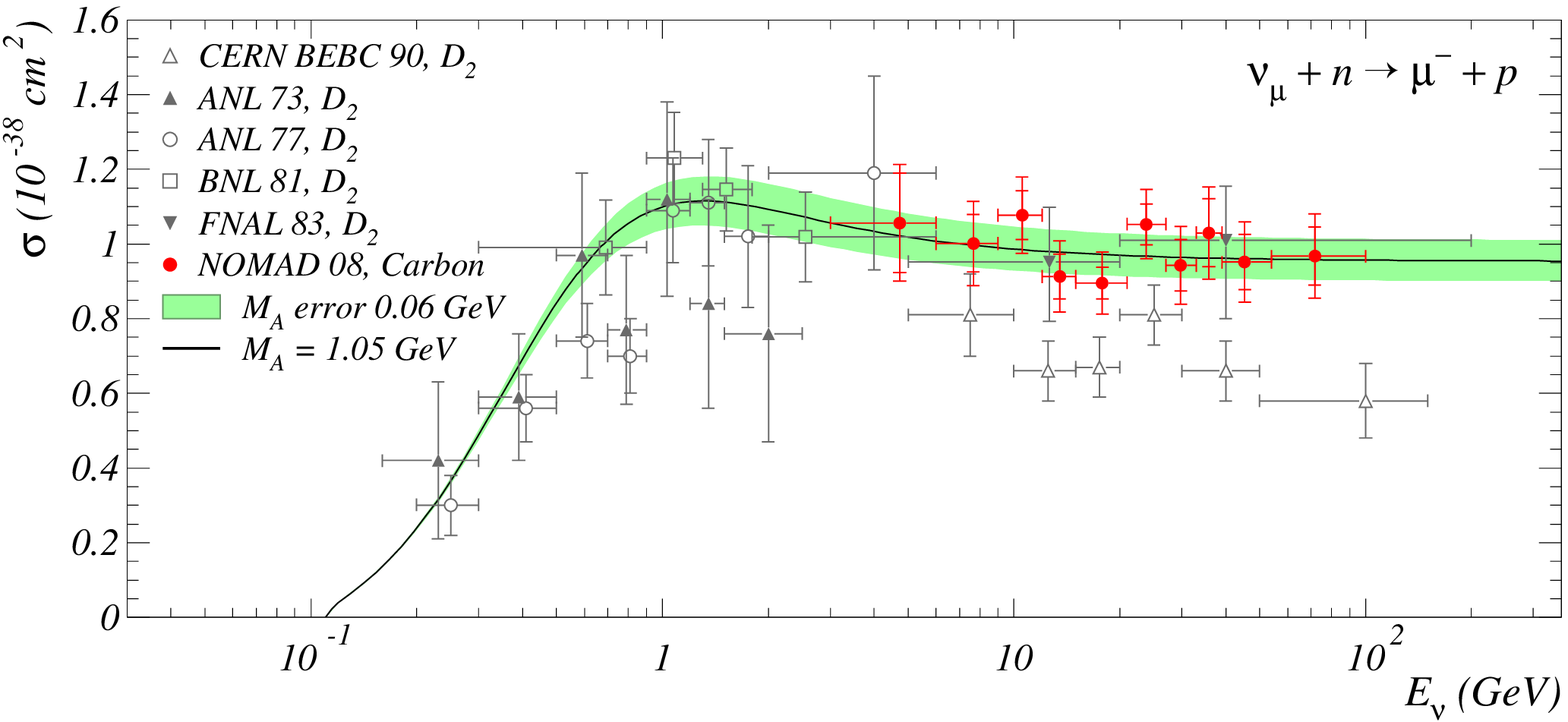,width=0.75\linewidth}}
\end{tabular}
\end{center}
\caption{\label{fig:nomad_cc_free}
Comparison of NOMAD $\cs{qel}_{\nu_\mu}$ measurements with previous $\nu_{\mu}D$ experimental data
(ANL~73 (Argonne 12-foot BC)~\cite{Mann:1973pr}, 
ANL~77~\cite{Barish:1977qk}, 
BNL~81 (Brookhaven 7-foot BC)~\cite{Baker:1981su}, 
FNAL~83 (FermiLab 15-foot BC)~\cite{Kitagaki:1983px}, 
BEBC~90 (CERN, Big European Bubble Chamber)~\cite{Allasia:1990uy}; 
corrections for nuclear effects have been made by the authors of the experiments).
The solid line and error band corresponds 
to the $M_A$ value obtained in the NOMAD experiment. 
}
\end{figure*}

\begin{figure*}
\begin{center}
\begin{tabular}{c}
\mbox{\psfig{file=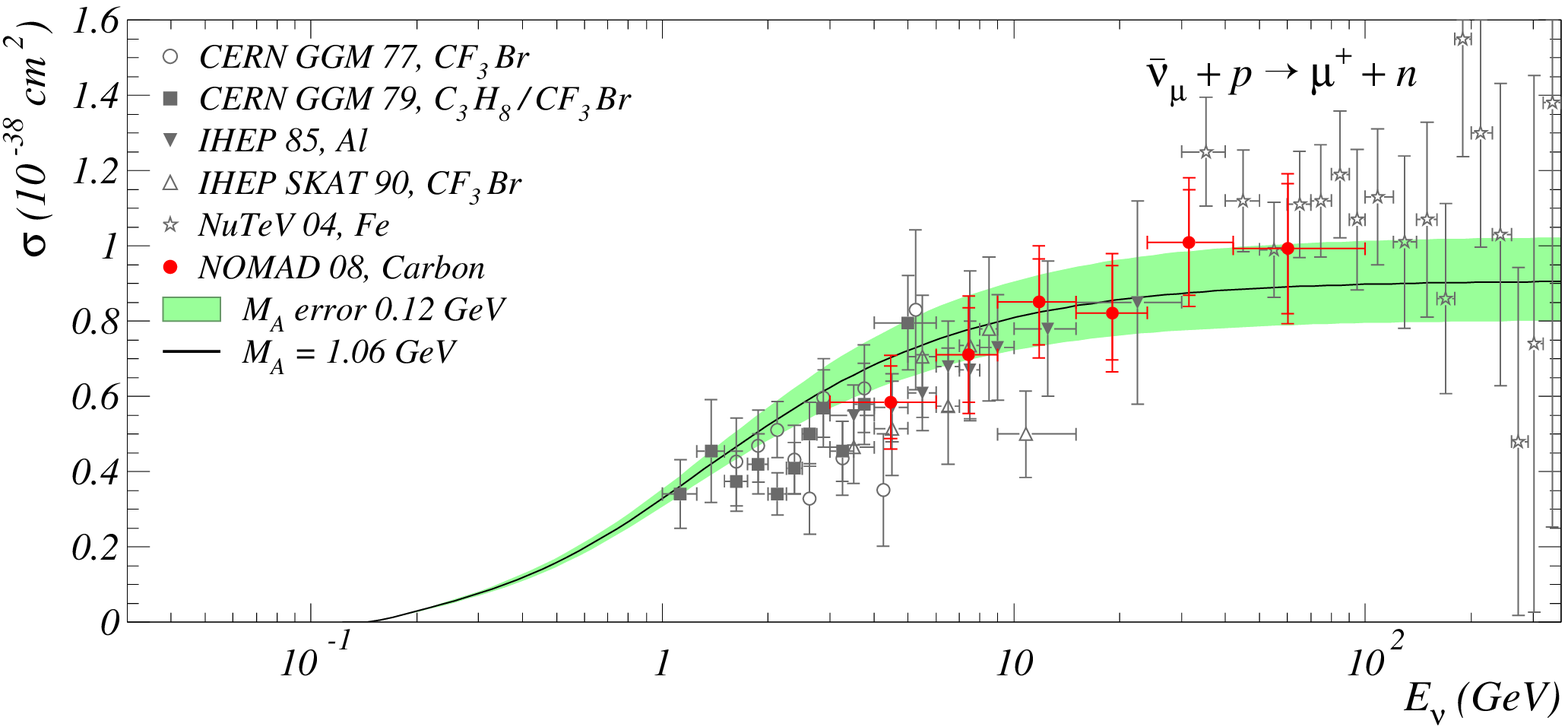,width=0.75\linewidth}}
\end{tabular}
\end{center}
\caption{\label{fig:nomad_ac_nucl}
Comparison of NOMAD $\cs{qel}_{\bar{\nu}_\mu}$ measurements, 
Table~\ref{table:nomad_qel_ac}, with previous experimental data 
on $\overline{\nu}_{\mu}$~scattering off heavy nuclei
(CERN~GGM~77~\cite{Bonetti:1977cs}, CERN~GGM~79~\cite{Armenise:1979zg},
IHEP~85~\cite{Belikov:1983kg}, IHEP~SCAT~90~\cite{Brunner:1989kw} and
NuTeV~04~\cite{Suwonjandee:04}).
The solid line and error band corresponds 
to the $M_A$ value obtained in the NOMAD experiment,
the error band takes into account both statistical and systematic uncertainties
of the present analysis. Nuclear effects 
are included into calculations according to the standard relativistic 
Fermi gas model. 
}
\end{figure*}

\subsection{$\nu_\mu n\to \mu^- p$ sample}

The results of our analysis for the $\nu_\mu$ sample are summarized in Table~\ref{table:results_cc}. We measure the flux averaged QEL cross-section in the neutrino energy 
interval $3-100~\GeV$ (see Eq.~(\ref{eq:cs_qel_i})) for the 1-track and 2-track samples 
as well as for their mixture (which is called {\em Combined} in 
Table~\ref{table:results_cc}). For each $\cs{qel}$ we calculate the corresponding
axial mass value, $M_A$. Results on $M_A$ extraction both from the 
standard $Q^2$ fit and from the combined $\tlD$ fit are also given.
These measurements are repeated for several QEL MC with different values of 
input parameters (the axial mass $M_A$ was varied between $0.83$ and $1.23~\GeV$  in steps of
$0.1~\GeV$; 
the formation time $\tau_0$ was allowed to take a value of 0.6, 1.0 and 2.0; 
the correction factor $\alpha_{mod}^F$ was varied within the  interval $[0.54,0.69]$). 
On top of this the NUANCE QEL MC with its own treatment of FSI effects is used for cross-checks.

We then observe that $M_A$ recalculated from the measured $\cs{qel}$ depends on 
$\tau_0$ if one refers to the 1-track or the 2-track samples. Specifically, the measured $M_A$ value 
increases with increasing $\tau_0$ when extracted from the 1-track sample 
while it decreases when extracted from the 2-track sample. 
This can be understood if we take into account the fact that the $\tau_0$ 
parameter controls the 
probability for an outgoing nucleon to be involved in an intranuclear cascade.
Increasing $\tau_0$ then increases the fraction of QEL events with 
reconstructed proton and thus populates the 2-track sample to the
detriment of the 1-track sample.
This is the reason for the systematic overestimation
of $M_A$ extracted from the 1-track sample alone and its underestimation
when extracted from the 2-track sample alone.
However the value of $M_A$ extracted from the combination of the
1-track and 2-track samples is almost insensitive to variations
of the $\tau_0$ parameter. 

We also find that using the QEL Monte Carlo with $\tau_0=1$ and $\alpha_{mod}^F=0.6$ provides the 
most accurate 
prediction for the ratio between the 1-track and 2-track
samples (and hence the most adequate description of the FSI):
in this case the flux averaged QEL cross-section 
stays approximately the same whether measured from the 1-track 
sample or from the 2-track sample (see Table~\ref{table:results_cc}). 
This allows us to exclude the MC sets with $\tau_0=0.6$ and $2.0$ from
further considerations.

Fig.~\ref{fig:nomad_cc_carbon_t} shows a comparison of our 
$\cs{qel}_{\nu_\mu}$ measurements as a function of
the neutrino energy in the 1-track and 2-track 
subsamples (for the best parameter $\tau_0 = 1.0$) with the final 
$\cs{qel}_{\nu_\mu}$ values measured using the full event sample.

Similarly we have observed that when using the full sample (1-track and 
2-track) the measured $M_A$ is not very sensitive to modifications of the
 $\alpha_{mod}^F$ parameter. And using the NUANCE simulation code as
a cross check gives a very consistent picture: the $M_A$ value extracted
from the 1-track sample is also different from the one extracted from
the 2-track  sample, while the value obtained with the combined sample
nicely agrees with our measurement with the best FSI parameters.
Thus, our results for the neutrino case are:
\begin{align}
 \label{eq:ma_recalculated}
 \cs{qel}_{\nu_\mu} & = (0.92 \pm 0.02 (stat) \pm 0.06 (syst))\times 10^{-38}~\cm^2
 \nonumber \\ 
 M_A & = 1.05 \pm 0.02 (stat) \pm 0.06 (syst)~\GeV
\end{align}
This result (\ref{eq:ma_recalculated}) is indeed in agreement with 
both the standard fit of the $Q^2$ distribution:
\begin{equation}
 \label{eq:ma_q2_fitted}
 M_A = 1.07 \pm 0.06 (stat) \pm 0.07 (syst)~\GeV
\end{equation}
and the fit of the combined $\tlD$ distribution of the NOMAD data:
\begin{equation}
 \label{eq:ma_fitted}
 M_A = 1.06 \pm 0.02 (stat) \pm 0.06 (syst)~\GeV
\end{equation}
(see Table~\ref{table:results_cc}, these results are obtained with 
a QEL MC using $M_A=1.03~\GeV$).

We use the 2-track sample only to extract $M_A$ from the fit of the $Q^2$ distribution
since in this case the purity of QEL identification is 
rather high ($\sim74\%$, see Table~\ref{tab:qel_identification}).
The results depend on the input
MC parameters (axial mass and formation time) but still are in nice agreement 
with the results of the extraction of $M_A$ from the measured QEL 
cross-section based also on a 2-track sample analysis. This can be considered 
as an additional confidence for our measurements using the full QEL sample.

The measured cross-section of the $\nu_\mu n\to \mu^- p$ reaction as a function of
the neutrino energy is presented in Table~\ref{table:nomad_qel_cc} and is shown in Figs.~\ref{fig:nomad_cc_nucl}
and~\ref{fig:nomad_cc_free}. These results are compared to the previous measurements
performed with deuterium and heavy nuclei targets 
(see discussion in Section~\ref{section:exp_data}).

\subsection{$\bar\nu_\mu p\to \mu^+ n$ sample}

In the $\bar\nu_\mu$ case the event topology is just a single $\mu^+$, 
thus the uncertainties in the treatment of FSI effect almost do not influence
the event selection.
Since our measurement of the cross-section 
of the $\bar\nu_\mu p\to \mu^+ n$ reaction is based
on a 1-track sample only,  
we do not show the dependence of the results on the
variation of the $\tau_0$ and $\alpha_{mod}^F$ parameters. Instead we display a dependence
on the input $M_A$ in Table~\ref{table:results_ac}. The results for the measured $M_A$ are
found to be quite stable. In Fig.~\ref{fig:nomad_ac_nucl} we show the measured
$\bar\nu_\mu p\to \mu^+ n$ cross section as a function of the antineutrino energy superimposed with the
theoretical curve drawn with $M_A=1.06\pm 0.12~\GeV$ and with nuclear effects
according to the standard relativistic Fermi gas model. 
Table~\ref{table:nomad_qel_ac} summarizes our results for the $\bar\nu_\mu p\to
\mu^+ n$ cross-section measurement in the different antineutrino
energy intervals. The cross-sections are measured on a Carbon target
and also recalculated for a free nucleon. The statistical and systematic errors are both
provided. The observed number of events in the data, the predicted number of
background events, the background subtracted and efficiency corrected number of events are also shown.

Our final results for the antineutrino case are:
\begin{align}
 \label{eq:ma_anumu}
 \cs{qel}_{\bar{\nu}_\mu} & = (0.81 \pm 0.05 (stat) \pm 0.08 (syst)) 
 \times 10^{-38}~\cm^2 \nonumber \\ 
 M_A & = 1.06 \pm 0.07 (stat) \pm 0.10 (syst)~\GeV
\end{align}

\begin{table*}
\caption{\label{table:nomad_qel_cc}
Cross-section of quasi-elastic neutrino scattering (in units of $10^{-38}\,\cm^2$,
statistical and systematic errors). $\sigma_{Carbon}$ is measured for NOMAD nuclear 
target and normalized per 1~neutron; $\sigma_{Free} = \sigma_{Nucl}/g$ is 
the cross-section for the free target neutron 
(the factor $g$ is calculated according to the Smith-Moniz model, 
see~\cite{Kuzmin:2007kr}). The number of 
selected events in raw data $N_{dat}$, the estimated background contamination $N_{bg}$ 
and the number of events $N_{cor}$ corrected for background and efficiency are also reported. The difference in the total number of data events with respect 
to Table~\ref{tab:qel_identification} (13683 vs 14021) is due to 
the additional cut on the neutrino energy $3 < E_\nu (\mbox{GeV}) < 100$. 
}
\begin{tabular}
{r@{ -- }r r >{\hspace*{12pt}}r r r >{\hspace*{18pt}}r r r >{\hspace*{18pt}}r r r}
\hline\noalign{\smallskip}
\multicolumn{2}{r}{$E_\nu$} & $\langle E_\nu \rangle$ & $N_{dat}$ & $N_{bg}$ & $N_{cor}$ &
\multicolumn{3}{r}{$(\sigma \pm \delta\sigma_{stat}  \pm \delta\sigma_{syst})_{Nucl}$} &
\multicolumn{3}{r}{$(\sigma \pm \delta\sigma_{stat}  \pm \delta\sigma_{syst})_{Free}$}  \\
\noalign{\smallskip}\hline\noalign{\smallskip}
   3 &    6 &    4.7 &    396 &   211.6 &   660.6 &   0.994 &   0.125 &   0.078 &   1.057 &   0.133 &   0.083 \\
   6 &    9 &    7.7 &   1115 &   580.2 &  1663.4 &   0.942 &   0.072 &   0.078 &   1.001 &   0.077 &   0.083 \\
   9 &   12 &   10.5 &   1683 &   835.3 &  2591.1 &   1.014 &   0.061 &   0.075 &   1.077 &   0.065 &   0.080 \\
  12 &   15 &   13.5 &   1647 &   834.9 &  2310.7 &   0.859 &   0.057 &   0.070 &   0.913 &   0.060 &   0.075 \\
  15 &   21 &   17.8 &   2815 &  1451.6 &  3766.8 &   0.843 &   0.040 &   0.067 &   0.896 &   0.043 &   0.071 \\
  21 &   27 &   23.8 &   2040 &   956.2 &  3084.7 &   0.991 &   0.052 &   0.070 &   1.053 &   0.055 &   0.075 \\
  27 &   33 &   29.8 &   1279 &   610.5 &  1816.8 &   0.888 &   0.064 &   0.073 &   0.943 &   0.068 &   0.077 \\
  33 &   39 &   35.8 &    852 &   400.9 &  1246.2 &   0.970 &   0.086 &   0.078 &   1.030 &   0.091 &   0.083 \\
  39 &   54 &   45.3 &   1008 &   496.1 &  1397.8 &   0.896 &   0.070 &   0.074 &   0.951 &   0.074 &   0.078 \\
  54 &  100 &   71.7 &    848 &   416.9 &  1191.5 &   0.911 &   0.073 &   0.077 &   0.967 &   0.078 &   0.082 \\
\noalign{\smallskip}\hline\noalign{\smallskip}
   3 &  100 &   23.4 &  13683 &  6794.2 & 19718.2 &   0.919 &   0.017 &   0.060 &   0.976 &   0.018 &   0.063 \\
\noalign{\smallskip}\hline
\end{tabular}
\end{table*}

\begin{table*}
\caption{\label{table:nomad_qel_ac}
The same as Table~\ref{table:nomad_qel_cc} but for antineutrino. \hfill
}
\begin{tabular}
{r@{ -- }r r >{\hspace*{12pt}}r r r >{\hspace*{18pt}}r r r >{\hspace*{18pt}}r r r}
\hline\noalign{\smallskip}
\multicolumn{2}{r}{$E_\nu$} & $\langle E_\nu \rangle$ & $N_{dat}$ & $N_{bg}$ & $N_{cor}$ &
\multicolumn{3}{r}{$(\sigma \pm \delta\sigma_{stat} \pm \delta\sigma_{syst})_{Nucl}$} &
\multicolumn{3}{r}{$(\sigma \pm \delta\sigma_{stat} \pm \delta\sigma_{syst})_{Free}$}  \\
\noalign{\smallskip}\hline\noalign{\smallskip}
   3 &    6 &    4.5 &    291 &   176.4 &   181.9 &   0.585 &   0.097 &   0.079 &   0.639 &   0.106 &   0.086 \\
   6 &    9 &    7.4 &    292 &   182.6 &   159.5 &   0.710 &   0.125 &   0.093 &   0.767 &   0.135 &   0.100 \\
   9 &   15 &   11.8 &    464 &   286.2 &   254.1 &   0.851 &   0.114 &   0.096 &   0.913 &   0.122 &   0.103 \\
  15 &   24 &   19.0 &    425 &   274.0 &   210.8 &   0.822 &   0.125 &   0.095 &   0.878 &   0.134 &   0.102 \\
  24 &   42 &   31.5 &    411 &   252.5 &   223.0 &   1.009 &   0.141 &   0.098 &   1.075 &   0.150 &   0.104 \\
  42 &  100 &   60.1 &    288 &   181.9 &   145.9 &   0.993 &   0.173 &   0.099 &   1.056 &   0.184 &   0.105 \\
\noalign{\smallskip}\hline\noalign{\smallskip}
   3 &  100 &   19.7 &   2171 &  1353.6 &  1182.5 &   0.811 &   0.053 &   0.081 &   0.866 &   0.056 &   0.086 \\
\noalign{\smallskip}\hline
\end{tabular}
\end{table*}

\begin{figure*}
\begin{center}
\mbox{\epsfig{file=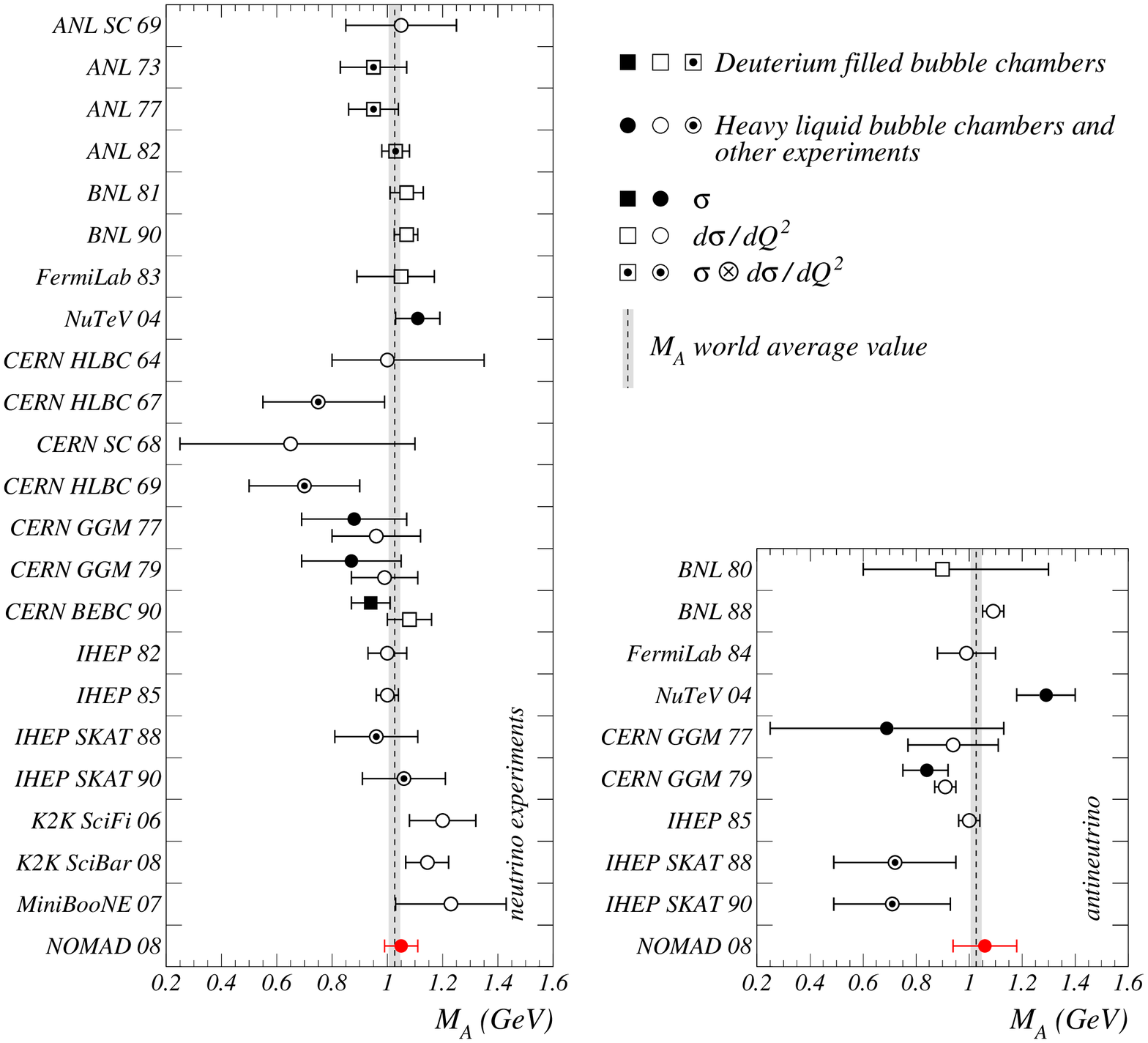,width=0.82\linewidth}}
\end{center}
\caption{\label{fig:ma_exp} A summary of existing experimental data: the axial 
  mass $M_A$ as measured in neutrino (left) and antineutrino (right)
  experiments. Points show results obtained both from deuterium filled BC (squares)
  and from heavy liquid BC and other experiments (circles). Dashed line corresponds to
  the so-called world average value $M_A = 1.026\pm 0.021~\GeV$ (see 
  review~\cite{Bernard:2001rs}). 
}
\end{figure*}

\section{\label{section:discussion} Conclusions}

The cross-section measurement of the $\nu_\mu n\to \mu^- p$ and 
$\bar\nu_\mu p\to \mu^+ n$ reactions on nuclear target was performed and 
reported in this article. 
The samples used in the analysis consist of 14021 neutrino and 2237 antineutrino events, which
were identified as quasi-elastic neutrino scattering among the experimental
data collected by the NOMAD collaboration.

We have discussed in details the analysis procedure and the most significant sources of systematic error. Special
attention was paid
to the influence of the FSI effects on the measured physical 
quantities. The DPMJET code was used to simulate these FSI effects.
We also proposed a method for tuning the intranuclear cascade
parameters (mainly the formation time $\tau_0$), which was then used to reduce the
corresponding systematic uncertainty. 

For the $\nu_\mu$ case stable results have been obtained
with the combined 1-track and 2-track samples 
since they are almost insensitive to the FSI effects.

The results for the flux averaged QEL cross-sections in 
the (anti)neutrino energy interval $3-100~\GeV$ are 
$\cs{qel}_{\nu_\mu} = (0.92 \pm 0.02 (stat) \pm 0.06 (syst))\times 10^{-38}~\cm^2$
and 
$\cs{qel}_{\bar{\nu}_\mu} = (0.81 \pm 0.05 (stat) \pm 0.08 (syst)) \times 10^{-38}~\cm^2$
for neutrino and antineutrino, respectively.

The axial mass $M_A$ was calculated from the measured cross-sections: we find 
$M_A = 1.05\pm 0.06~\GeV$ from the $\nu_\mu$ sample and $M_A = 1.06\pm 0.12~\GeV$ from
the $\bar\nu_\mu$ sample. The $M_A$ parameter was also extracted from the fit of the $Q^2$
distribution in the high purity sample of $\nu_\mu$ quasi-elastic 2-track events (with
a reconstructed proton track). It was found to be consistent with the values
calculated from the cross-sections. 

Our results are in agreement with the existing world average 
value~\cite{Bernard:2001rs,Bodek:2007ym} 
and do not support the results found in recent
measurements from the NuTeV~\cite{Suwonjandee:04}, K2K~\cite{Gran:2006jn,Mariani:2008zz} and
MiniBooNE~\cite{MiniBooNE:2007ru} collaborations, which reported somewhat larger
values, however still compatible with our results within their large errors.
A summary of existing experimental data on the axial 
mass measurements in neutrino and antineutrino
experiments is shown in Fig.~\ref{fig:ma_exp}.

It should also be noted that the preliminary results reported earlier 
by the NOMAD collaboration for 
the 2-track  sample only~\cite{Petti:2004wy,Lyubushkin:2006dv} 
suffered from a large systematic bias related 
to an improper treatment of the FSI effects in the simulation program. 
They should be now superseeded by the new measurements reported here.

\section{Acknowledgements}

The experiment was supported by the following agencies:
Australian Research Council (ARC) and Department of Industry, Science, and
Resources (DISR), Australia;
Institut National de Physique Nucl\'eaire et Physique des Particules (IN2P3),
Commissariat \`a l'Energie Atomique (CEA), Minist\`ere de l'Education
Nationale, de l'Enseigne\-ment Sup\'erieur et de la Recherche, France;
Bundesministerium f\"ur Bildung und Forschung (BMBF), Germany;
Istituto Nazionale di Fisica Nucleare (INFN), Italy;
Institute for Nuclear Research of the Russian Academy of Sciences, Russia;
Joint Institute for Nuclear Research;
Russian Foundation for Basic Research (grant 08-02-00018);
Fonds National Suisse de la Recherche Scientifique, Switzerland;
Department of Energy, National Science Foundation,
the Sloan and the Cottrell Foundations, USA.

We thank the management and staff of CERN and of all participating institutes 
for their vigorous support of the experiment. Particular thanks are due to the 
CERN SPS accelerator and beam-line staff for the magnificent performance of 
the neutrino beam. Special thanks for useful discussions of theoretical issues 
go to K.~Kuzmin and V.~Naumov; we are grateful to prof. J.~Ranft for the 
important explanations and technical assistance with the DPMJET code. V.~Lyubushkin 
is very grateful to the LPNHE (Paris) for the warm hospitality and financial support 
during the final stage of this work.

\begin{table*}
\caption{\label{table:results_cc}
Parameters of the QEL MC simulation (axial mass $M_A^{mc}$ and parameters of FSI
modeling) are listed in the first three columns. The intermediate columns contain
results of the QEL $\nu_\mu$ cross-section measurement (in units of $10^{-38}\,\cm^2$,
without errors) for the different topology of identified events (with or without
reconstructed proton track). The axial mass value obtained from the fit of $Q^2$
distribution and $\tlD$ fit ($Q^2_{lim} = 0.2-4 \,\rm{GeV}^2$) are given in the last 
columns of the table; the given statistical errors on the axial mass $\delta M_A$ are from MINUIT output.
\hfill
}
\begin{tabular}
{r r r >{\hspace*{5pt}}rr >{\hspace*{5pt}}rr >{\hspace*{5pt}}rr >{\hspace*{5pt}}rrr rrr}
\hline\noalign{\smallskip}
\multicolumn{3}{r}{\em MC parameters} & 
\multicolumn{2}{r}{\em Single track} & \multicolumn{2}{r}{\em Two tracks} & \multicolumn{2}{r}{\em Combined} &
\multicolumn{3}{r}{\em Fit of $Q^2$ distribution} &
\multicolumn{3}{r}{\em Fit of $\tlD$} \\
$\tau_0$ & $\alpha_{mod}^F$ & $M_A^{mc}$ &
$\sigma_{qel}$ & $M_A$ & $\sigma_{qel}$ & $M_A$ & $\sigma_{qel}$ & $M_A$ &
\multicolumn{2}{r}{$M_A \pm \delta M_A$} & $\chi^2$ &
\multicolumn{2}{r}{$M_A \pm \delta M_A$} & $\chi^2$ \\
\noalign{\smallskip}\hline\noalign{\smallskip}
    0.60 &   0.60 &   0.83 & 0.863 & 0.990 & 1.014 & 1.148 & 0.915 & 1.047 &  1.113 & 0.057  & 19.4 &  1.105 &  0.018 & 19.3 \\  
    1.00 &   0.60 &   0.83 & 0.885 & 1.015 & 0.956 & 1.090 & 0.912 & 1.043 &  1.095 & 0.057  & 11.6 &  1.058 &  0.018 & 11.9 \\  
    2.00 &   0.60 &   0.83 & 0.918 & 1.050 & 0.851 & 0.977 & 0.892 & 1.021 &  0.960 & 0.093  & 17.4 &  0.965 &  0.018 & 17.4 \\  
\noalign{\smallskip}\hline\noalign{\smallskip}
    0.60 &   0.60 &   0.93 & 0.882 & 1.011 & 1.015 & 1.148 & 0.928 & 1.061 &  1.135 & 0.056  & 10.0 &  1.119 &  0.018 & 10.0 \\  
    1.00 &   0.60 &   0.93 & 0.893 & 1.023 & 0.942 & 1.074 & 0.911 & 1.043 &  1.075 & 0.060  & 13.5 &  1.065 &  0.018 & 13.5 \\  
    2.00 &   0.60 &   0.93 & 0.931 & 1.063 & 0.844 & 0.968 & 0.896 & 1.026 &  1.009 & 0.069  & 12.2 &  0.971 &  0.018 & 12.4 \\  
\noalign{\smallskip}\hline\noalign{\smallskip}
    0.60 &   0.60 &   1.03 & 0.910 & 1.041 & 0.977 & 1.110 & 0.935 & 1.067 &  1.016 & 0.051  & 25.8 &  1.094 &  0.017 & 27.1 \\  
    1.00 &   0.60 &   1.03 & 0.919 & 1.051 & 0.918 & 1.050 & 0.919 & 1.051 &  1.073 & 0.059  & 18.7 &  1.059 &  0.018 & 18.7 \\  
    2.00 &   0.60 &   1.03 & 0.950 & 1.083 & 0.819 & 0.939 & 0.896 & 1.026 &  0.993 & 0.079  & 18.4 &  0.968 &  0.018 & 18.4 \\  
\noalign{\smallskip}\hline\noalign{\smallskip}
    0.60 &   0.60 &   1.13 & 0.946 & 1.079 & 0.979 & 1.113 & 0.959 & 1.092 &  1.031 & 0.077  & 24.9 &  1.109 &  0.017 & 26.4 \\  
    0.80 &   0.60 &   1.13 & 0.948 & 1.081 & 0.926 & 1.058 & 0.940 & 1.073 &  1.092 & 0.056  & 13.8 &  1.079 &  0.018 & 13.8 \\  
    1.00 &   0.60 &   1.13 & 0.962 & 1.096 & 0.904 & 1.035 & 0.940 & 1.072 &  1.100 & 0.062  & 19.4 &  1.060 &  0.018 & 19.6 \\  
    2.00 &   0.60 &   1.13 & 0.995 & 1.129 & 0.789 & 0.904 & 0.906 & 1.037 &  0.999 & 0.080  & 18.0 &  0.956 &  0.018 & 18.2 \\  
\noalign{\smallskip}\hline\noalign{\smallskip}
    0.60 &   0.60 &   1.23 & 0.994 & 1.127 & 0.925 & 1.058 & 0.967 & 1.100 &  1.039 & 0.053  & 20.7 &  1.088 &  0.018 & 21.2 \\  
    0.80 &   0.60 &   1.23 & 0.996 & 1.129 & 0.904 & 1.035 & 0.959 & 1.092 &  1.013 & 0.039  & 21.1 &  1.066 &  0.017 & 21.7 \\  
    1.00 &   0.60 &   1.23 & 1.000 & 1.134 & 0.879 & 1.008 & 0.951 & 1.085 &  0.970 & 0.087  & 20.1 &  1.051 &  0.017 & 21.3 \\  
    2.00 &   0.60 &   1.23 & 1.038 & 1.171 & 0.777 & 0.889 & 0.921 & 1.053 &  0.996 & 0.079  & 20.9 &  0.964 &  0.018 & 16.3 \\  
\noalign{\smallskip}\hline\noalign{\smallskip}
    0.80 &   0.54 &   1.03 & 0.921 & 1.053 & 0.963 & 1.097 & 0.937 & 1.070 &  1.113 & 0.054  & 20.9 &  1.079 &  0.018 & 21.1 \\  
    0.80 &   0.57 &   1.03 & 0.921 & 1.052 & 0.950 & 1.083 & 0.932 & 1.064 &  1.072 & 0.062  & 15.5 &  1.067 &  0.018 & 15.5 \\  
    0.80 &   0.60 &   1.03 & 0.920 & 1.051 & 0.959 & 1.092 & 0.935 & 1.067 &  1.090 & 0.064  & 12.7 &  1.089 &  0.018 & 12.7 \\  
    0.80 &   0.63 &   1.03 & 0.912 & 1.044 & 0.953 & 1.087 & 0.928 & 1.060 &  1.082 & 0.062  & 15.8 &  1.084 &  0.018 & 15.7 \\  
    0.80 &   0.66 &   1.03 & 0.905 & 1.035 & 0.933 & 1.066 & 0.916 & 1.047 &  0.989 & 0.091  & 19.9 &  1.067 &  0.017 & 20.9 \\  
    0.80 &   0.69 &   1.03 & 0.904 & 1.035 & 0.940 & 1.072 & 0.918 & 1.049 &  0.937 & 0.113  & 15.7 &  1.070 &  0.017 & 18.0 \\  
\noalign{\smallskip}\hline
\end{tabular}
\end{table*}

\begin{table*}
\caption{\label{table:results_ac}
  The results of QEL $\bar{\nu}_\mu$ cross-section measurement. 
  The parameters of the DPMJET model are 
  $\tau_0 = 1.0$, $\alpha_{mod}^F = 0.6$. 
  \hfill
}
\begin{tabular}
{r >{\hspace*{12pt}}r >{\hspace*{6pt}}r}
\hline\noalign{\smallskip}
{$M_A^{mc}$} & {$\sigma_{qel}$} & {$M_A$} \\
\noalign{\smallskip}\hline\noalign{\smallskip}
0.83 & 0.794 & 1.042  \\  
0.93 & 0.799 & 1.048  \\  
1.03 & 0.811 & 1.063  \\  
1.13 & 0.834 & 1.094  \\  
1.23 & 0.861 & 1.127  \\  
\noalign{\smallskip}\hline
\end{tabular}
\end{table*}



%
%

\end{document}